\documentclass[journal]{IEEEtran}

\usepackage{ifpdf}

\usepackage{cite}

\ifCLASSINFOpdf
   \usepackage[pdftex]{graphicx}
\else
\fi
\usepackage{amsmath}
\usepackage{amssymb}
\usepackage{bm}
\DeclareMathOperator*{\mean}{mean}
\DeclareMathOperator*{\size}{size}
\DeclareMathOperator*{\vectorize}{vec}
\DeclareMathOperator*{\diag}{diag}
\usepackage{algorithmic}

\usepackage{array}
\renewcommand{\tablename}{TABLE~}

\newtheorem{corollary}{\bf Corollary}
\newtheorem{theorem}{\bf Theorem}

\newtheorem{lemma}{\bf Lemma}

\begin{document}

%
\title{On Clutter Ranks of Frequency Diverse Radar Waveforms}
%
%
%

\author{
Yimin~Liu,~\IEEEmembership{Member,~IEEE,}
	Le~Xiao,
	Xiqin~Wang,
	and Arye~Nehorai,~\IEEEmembership{Fellow,~IEEE}
\thanks{Y. Liu, L. Xiao, and X. Wang are with the Department of Electronic Engineering, Tsinghua University, Beijing, 100084, China. e-mail: yiminliu@tsinghua.edu.cn, xiaol13@mails.tsinghua.edu.cn, wangxq\_ee@tsinghua.edu.cn.}
\thanks{A. Nehorai is with the Department of Electrical and Systems Engineering, Washington University, St. Louis, MO 63130 USA. e-mail: nehorai@ese.wustl.edu.}
\thanks{The work of Y. Liu was supported by the National Natural Science Foundation of China (Grant No. 61571260 and 61201356). The work of A. Nehorai was supported by the AFSOR (Grant No. FA9550-11-1-0210). Corresponding e-mail: yiminliu@tsinghua.edu.cn.}
}
\maketitle

\begin{abstract}
Frequency diverse (FD) radar waveforms are attractive in radar research and practice. By combining two typical FD waveforms, the frequency diverse array (FDA) and the stepped-frequency (SF) pulse train, we propose a general FD waveform model, termed the random frequency diverse multi-input-multi-output (RFD-MIMO) in this paper. The new model can be applied to specific FD waveforms by adapting parameters. Furthermore, by exploring the characteristics of the clutter covariance matrix, we provide an approach to evaluate the clutter rank of the RFD-MIMO radar, which can be adopted as a quantitive metric for the clutter suppression potentials of FD waveforms. Numerical simulations show the effectiveness of the clutter rank estimation method, and reveal helpful results for comparing the clutter suppression performance of different FD waveforms.
\end{abstract}


%
\IEEEpeerreviewmaketitle

\section{Introduction}
\IEEEPARstart{W}{aveform} diversity has led to many interesting and promising concepts in the research and practice of the radar community in the past decade. By exploring waveform adaptivity in different domains, such as the spatial (antenna beampattern), temporal, spectral, code, and polarization domains, remarkable improvements have been realized in radar abilities, such as high resolution imaging, target recognition, clutter suppression, and electronic-counter-countermeasures (ECCM) \cite{gini2012waveform}. Among the different kinds of diverse waveforms, frequency diverse (FD) waveforms are attractive due to their ease of use in system implementation \cite{Skolnik2002Introduction}, efficiency in wide-bandwidth synthesis \cite{wehner1995high}, and robust in spectral compatibility and resilience \cite{dahm2010technology}.

In 2006, a new array antenna, named the frequency diverse array (FDA), was introduced in \cite{Antonik2006Frequency}. By linearly \cite{Antonik2006Frequency} or randomly \cite{yimin2016ICASSP} (called LFDA or RFDA, respectively) assigning the carrier frequencies of array elements, an FDA can provide a beampattern which depends on both direction and range, and brings important benefits like transmit beamforming \cite{wang2013range}, target range-direction estimation \cite{wang2014range}, and jamming resistance \cite{xu2015space}, to list a few. Moreover, FDA-based algorithms enable advantages in clutter or interference discrimination, and in moving target detection. By using an FDA, the clutter in forward-looking radar was alleviated \cite{baizert2006forward}. In \cite{xu2015range}, clutter whose delay was outside of one pulse repetition interval (PRI) was successfully discriminated, hence the target detection performance of an airborne multi-input-multi-output (MIMO) radar was improved. However, quantitive metrics of FDA radars' clutter suppression performance are still inadequate.
\begin{figure*}[!t]
\centering
\includegraphics[width=.6\textwidth]{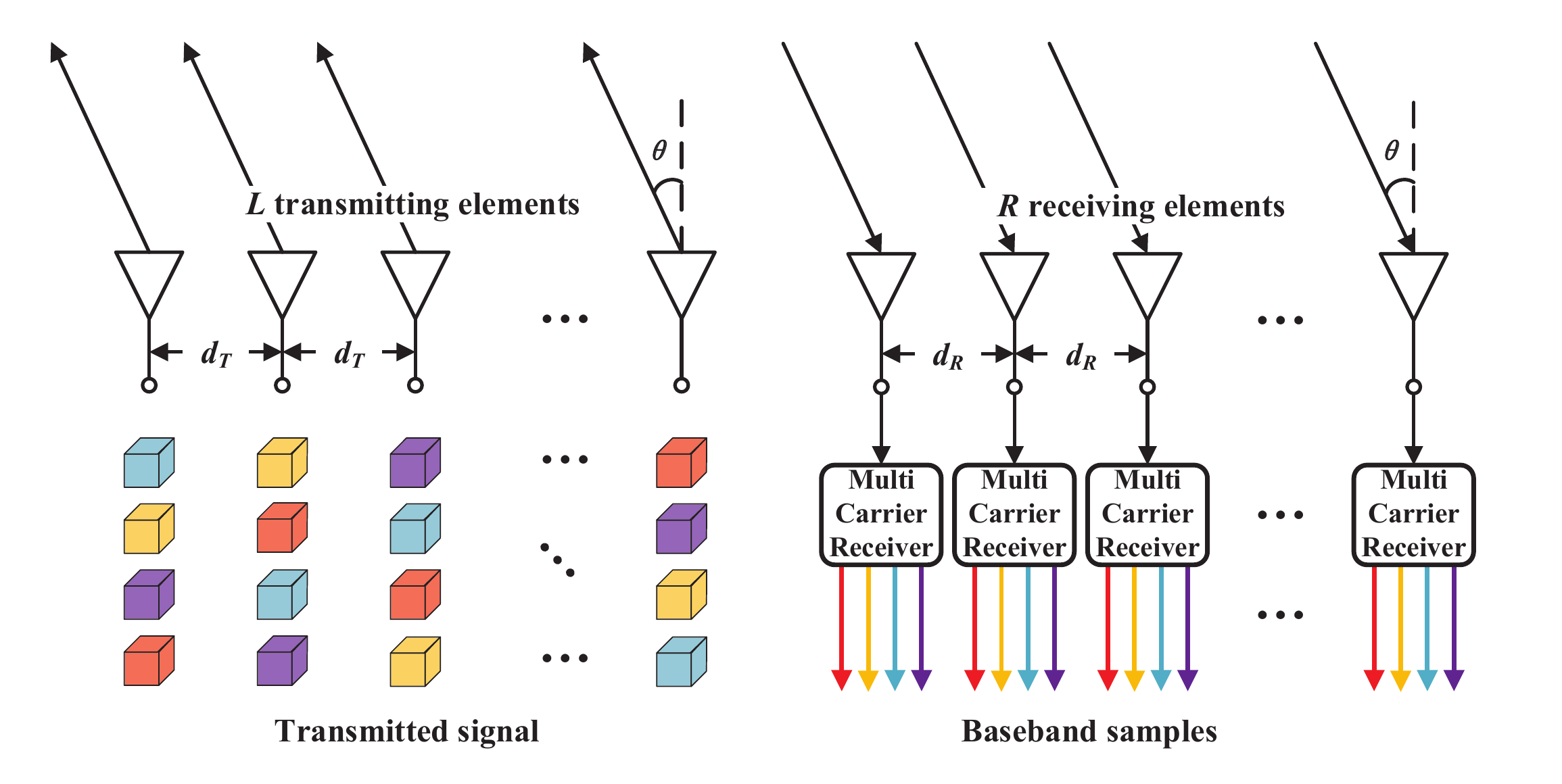}
\caption{A brief schematic of an RFD-MIMO radar. Each brick corresponds to a pulse in the waveform, and different colors signify different carrier frequencies. The rows and columns of the bricks represent the pulses and transmitting elements, respectively. }
\label{Fig:SystemSketch}
\end{figure*}

As defined in the IEEE Radar Standard P686/D2 (January 2008), frequency diversity radar is ``a radar that operates at more than one frequency, using either parallel
channels or sequential groups of pulses". Following this definition, the FDA can be regarded as one kind of FD waveform that is distributed in the spatial-spectral domain. Similarly, another kind of widely-used waveform, stepped-frequency (SF) pulse train \cite{levanon2002stepped}, can be regarded as FD waveforms distributed in the temporal-spectral domain. In the SF pulse train, the carrier frequencies of pulses in a coherent processing interval (CPI) shift linearly \cite{einstein1984generation} or randomly \cite{axelsson2007analysis} (termed linear SF (LSF), or random SF (RSF)). Basic LSF pulse train can synthesize a large bandwidth to achieve a very high range resolution \cite{wehner1995high}. Furthermore, if the carrier frequencies of successive pulses shift randomly, as in the RSF pulse trains, the ambiguity functions will become thumbtack-like, which implies an uncoupled high resolution in both range and velocity \cite{wehner1995high}. Moreover, the ECCM performances of RSF radars are also outstanding, due to its frequency agility \cite{Skolnik2002Introduction}.

Although some algorithms have been proposed for clutter suppression \cite{gill1995step}\cite{gaikwad2011application}, the corresponding performance evaluations for the SF, especially RSF radars, still need research. In this paper, we try to provide general quantitive information about the clutter suppression potential of the FD radar waveforms. A new FD waveform model, named the random-frequency-diverse-MIMO (RFD-MIMO), is proposed by integrating the FDA and SF pulse train. Because the carrier frequencies of  the array elements and pulses in RFD-MIMO vary agilely, the new waveform is diverse in a 3D (spatial-temporal-spectral) domain. Moreover, this model can be applied to existing FD waveforms by adapting the model parameters. Furthermore, the clutter rank, defined as the rank of a radar's clutter covariance matrix (CCM), is evaluated based on this model. As introduced in \cite{klemm2002principles} and \cite{goodman2007clutter}, the clutter rank can quantify the averaged clutter suppression performance of a radar. Hence the results of this paper can be regarded as a quantitive metric of the clutter suppression potential that can be achieved by the coherent processing of a frequency diversity radar. It should be noted that this study focuses on coherent approaches to clutter suppression, and the intention is different from conventional works which employed the de-correlated radar cross section (RCS) response properties. In addition, unlike traditional clutter rank studies, such as \cite{goodman2007clutter} and \cite{wang2010clutter}, the main challenge encountered by this work was the disordered phase relationships between array elements and pulses, caused by the frequency diversity. The main contributions of this paper are as follows.

\begin{enumerate}
\item
We propose a general model, named RFD-MIMO, of FD radar waveforms. In this model, the carrier frequency of each transmitting array element and each pulse can be assigned an arbitrary value, and each receiving array element can receive all the possible carrier frequencies simultaneously.
\item
We construct the target and clutter model of the RFD-MIMO waveform. Based on this model, we derive the expressions of the FDA, SF, and FD-MIMO radars' clutter.
\item
By exploring the features of an RFD-MIMO's CCM, we derive an approximation of the new FD waveform's clutter rank. We find that the frequency diversity radar's CCM is sparse, and can be permuted to a block diagonal matrix, which notably reduces the complexity of the clutter rank estimation. 
\item
We substantiate the clutter rank estimation of RFD-MIMO to specific FD radar waveforms, and quantify the clutter suppression potentials of different frequency diversity radars. The results reveal that, first, in radars using FDA or SF pulse trains, random carrier frequency assignments have advantages in clutter suppression over their linear counterparts. Second, wideband pulses and MIMO antennas are more suitable for target detection in heavy clutter scenarios.
\end{enumerate}

The rest of this paper is organized as follows. Section \ref{sec:SignalModel} presents the radar schematic, and constructs the system and signal models of the RFD-MIMO waveform. In Section \ref{Sec:ClutterRank}, by exploring the CCM, we derive an estimation approach for the clutter rank of the new waveform. In Section \ref{Sec:Extension}, discussion and numerical results for radars with specific FD waveforms are provided as substantiations of the provided method. Conclusions are drawn in the last section.

\textit{Notations}: The important and frequently used notations are listed in Table \ref{tab:Notations}.
\begin{table}[!t]
\centering
\caption{Glossary of Notations}
\label{tab:Notations}
\begin{tabular}{l|l}
\hline
  \textbf{Notation} & \textbf{Discription} \\
  \hline
$c$ & The speed of light\\

$t$ & The time variable\\

$\mathbb{Z}$ & The set of all integers\\

$\mathbb{C}$ & The set of all complex numbers\\

$f_c$& The initial carrier frequency\\

$\Delta{f}$ & The frequency increment\\

$d_T$ & The distance between transmitting antenna elements\\

$d_R$ & The distance between receiving antenna elements\\

$T$ & The pulse repetition interval (PRI)\\

$L$ & The number of transmitting antenna elements\\

$R$ & The number of receiving antenna elements\\

$P$ & The number of pulses in a pulse train\\

$\mathbf{G}$ & The frequency diverse code matrix\\

$\mathbf{G}_Q$ & The augmented frequency diverse code matrix\\
& with pulse bandwidth $Q\Delta{f}$\\

$\mathbb{M}_Q$ & The set composed of all the unique entries in $\mathbf{G}_Q$\\

$\mathbf{l}$ & The transmitting array element index vector\\

$\mathbf{r}$ & The receiving array element index vector\\

$\mathbf{p}$ & The pulse index vector\\

$\mathbf{q}$ & The sub-band index vector\\

$\mathbb{D}$ & The clutter range region\\

$\mathbb{V}$ & The clutter velocity region\\

$\mathbb{A}$ & The clutter direction sine region\\

$\mathcal{C}$ & The clutter rank\\

$\mathcal{L}_{\mathrm{FD}}$ & The frequency diversity loss (FDL)\\

$\mathcal{R}\{\cdot\}$ & The rank of a matrix\\

$\vectorize\{\cdot\}$ & Column vectorization of a matrix\\

$[\cdot]_{a, b}$ & The $a$th row, $b$th column entry of a matrix\\

$[\cdot]_{a}$ & The $a$th entry(column) of a vector(matrix)\\

$|\cdot|$ & The number of unique elements of a vector/matrix/set\\

$|\cdot|_2$ & The $l_2$-norm of a scalar/vector\\

$\langle\cdot\rangle$&The difference between the maximum and minimum \\
& entries in a vector/matrix/set\\

$\lfloor\cdot\rfloor$ & The largest integer which is no larger than an argument\\

$\lceil\cdot\rceil$ & The smallest integer which is no less than an argument\\

$(\cdot)^{T}$, $(\cdot)^H$ & The transpose and Hermitian of an argument\\

$(\cdot)^*$&  The element-wise complex conjugation of an argument\\

$\mathcal{I}_m^{\parallel}\{\cdot\}$ & The column vector composed of row indices corresponding\\
& to a vector/matrix's entries which equal to $m$\\

$\mathcal{I}_m^{=}\{\cdot\}$ & The column vector composed of column indices \\
& corresponding to a vector/matrix's entries which equal to $m$\\

$\mathbf{1}^{A\times B}$ & An $A\times B$ matrix (vector) with all-one entries\\

$\otimes$, $\odot$ & The Kronecker and Hadamard products\\

$\circledast$ & The Khrati-Rao product\\

$\oplus$ & The \textit{stretched} sum (defined in Section \ref{sec:SignalModel})\\

  \hline
\end{tabular}
\end{table}

\section{System and Signal Models}
\label{sec:SignalModel}

In this section, we first formulate the RFD-MIMO radar waveform as a general model of FD waveforms, and then give the expression of the echoes from targets and clutter for the RFD-MIMO and specific frequency diversity radars.
\subsection{General Model}
\label{subsec:SignalModel:GeneralModel}
As introduced in \cite{yimin2016ICASSP}, the FDA can be regarded as an FD waveform which is distributed in the spatial-spectral domain. Beyond this, we introduce a pulse train into the waveform to expand it from 2D (spatial-spectral) diversity to 3D (spatial-temporal-spectral) diversity. The system schematic of a radar with RFD-MIMO waveform is shown in Fig. \ref{Fig:SystemSketch}. In the RFD-MIMO radar, the transmitting and receiving antennas are colocated. There are $L$ array elements, indexed by $l=0,1,\dots, L-1$, in the transmitting antenna, and $R$ array elements, indexed by $r=0,1,\dots, R-1$, in the receiving antenna. The elements in both antennas are equally separated, with the inter-element distances of the transmitting and receiving antennas being $d_T$ and $d_R$, respectively.

In the RFD-MIMO radar, every transmitting element can be assigned an arbitrary carrier frequency which is chosen from a candidate frequency set. The carrier frequency of every element can (but is not necessarily required to) vary from pulse to pulse. Thus a carrier frequency in the waveform is equal to an initial frequency ($f_c$) plus an integral multiple of a frequency increment ($\Delta{f}$). The integer is named the frequency diverse code (FDC), and for all the $P$ pulses (indexed by $p=0,1,\dots, P-1$) in the pulse train, there are $PL$ integers, which can be arranged into a frequency diverse code matrix (FDCM), $\mathbf{G}\in\mathbb{Z}^{P\times L}$. Then the carrier frequency transmitted by the $l$th array element in the $p$th pulse is
\begin{equation}
\label{Eq:TransFrequency}
f_{p,l}=f_c+\Delta{f}[\mathbf{G}]_{p,l}.
\end{equation}

For the signal of each pulse, both narrowband and wideband cases are considered. In narrowband cases, each pulse is assumed as monotone, with the same frequency as its carrier frequency. In wideband cases, following the convention in \cite{sammartino2013frequency}, it is supposed that all the pulses have the same bandwidth $B$, which can be divided into $Q$ sub-bands ($Q$ is an integer, and $B=Q\Delta{f}$); the signal of every sub-band can be regarded as a monotone multiplied by a modulation coefficient.

Then the transmitted signal of the $p$th pulse, $l$th array element, and $q$th ($q=0,1,\dots Q-1$) sub-band is
\begin{equation}
\label{Eq:TransSubband}
s_{p,l,q}(t)=\beta_{p,l,q}\cdot\exp\Big(j2\pi\big(f_c+\Delta{f}([\mathbf{G}]_{p,l}+q)\big)t\Big),
\end{equation}
where $\beta_{p,l,q}$ is the modulation coefficient. According to (\ref{Eq:TransSubband}), the FDCM can be expanded to an augmented-FDCM (a-FDCM), given by
\begin{equation}
\label{Eq:AFDCM}
\mathbf{G}_Q = \mathbf{G}\oplus\mathbf{q},
\end{equation}
where $\mathbf{q}=[0,1,\dots, Q-1]^T$. In (\ref{Eq:AFDCM}), $\oplus$ is the \textit{stretched} sum operator, where $\mathbf{A}\oplus\mathbf{B}=\mathbf{A}\otimes\mathbf{1}^{\size(\mathbf{B})}+\mathbf{1}^{\size(\mathbf{A})}\otimes\mathbf{B}$ \footnote{The definition of $\size(\cdot)$ follows the eponymous MATLAB$^{\circledR}$ routine which returns the size of a matrix.}.

Then the RFD-MIMO waveform has $|\mathbb{M}_Q|$ frequency points, and (\ref{Eq:TransSubband}) can be rewritten as
\begin{equation}
\label{Eq:TransSubband2}
s_{p,l,q}(t)=\beta_{p,l,q}\cdot\exp\big(j2\pi(f_c+\Delta{f}[\mathbf{G}_Q]_{pQ+q,l})t\big).
\end{equation}

In addition, due to the colocated assignment of the MIMO antenna, the direction\footnote{For conciseness, we call both $\theta$ and $\alpha=\sin\theta$ ``direction", because they can be easily distinguished within their context.} $\alpha=\sin\theta$ and radial velocity $v$ of a point scatterer can be regarded as identical with respect to (w.r.t.) every transmitting and receiving array element \cite{li2009mimo}. Therefore, at the $p$th pulse, the time delay from the $l$th transmitting element to the scatterer and back to the $r$th receiving element is
\begin{equation}
\label{Eq:TimeDelay}
\tau_{p,l,r}(D,v,\alpha)=\frac{1}{c}(2D+2vpT+\alpha ld_T+\alpha rd_R),
\end{equation}
where $T$ is the pulse repetition interval (PRI), and $D$ is the initial range (when $t=0$) between the scatterer and the $0$th array element.
\begin{figure}[!t]
\centering
\includegraphics[width=.5\textwidth]{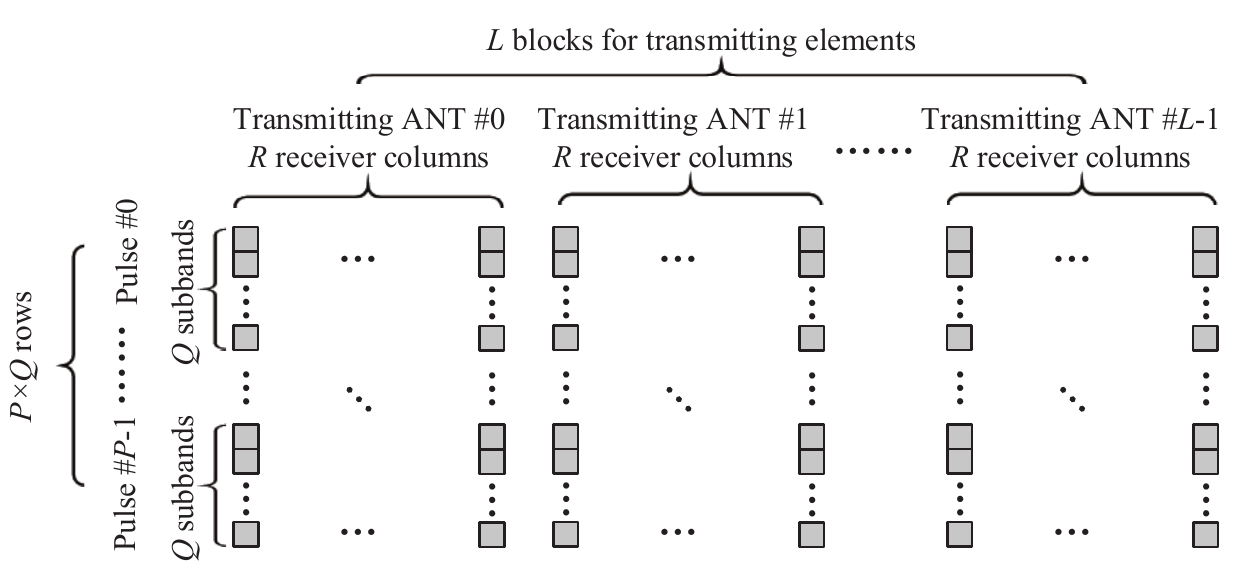}
\caption{An illustration of the data arrangement format of an RFD-MIMO waveform. The data matrix contains $PQ$ rows, where the $(pQ+q)$th row corresponds to the $q$th sub-band in the $p$th pulse. There are $LR$ columns, where the $(lR+r)$th column corresponds to the echo from the $l$th transmitting to the $r$th receiving array element.}
\label{Fig:DataMTX}
\end{figure}

Disregarding both the energy divergence in the wave propagation paths and the variation of the target's reflection factor, the received echo from a unit point scatterer can be seen as the time-delayed version of the transmitted signal. Thus the echo of the $q$th sub-band from the $l$th transmitting to the $r$th receiving array element is 
\begin{eqnarray}
\label{Eq:ReceivedSignal}
r_{p,l,r,q}(t;D,v,\alpha)&=&\beta_{p,l,q}\cdot\exp\Big(j2\pi(f_c+\Delta{f}[\mathbf{G}_Q]_{pQ+q,l})\nonumber\\
&&\cdot\big(t-\tau_{p,l,r}(D,v,\alpha)\big)\Big).
\end{eqnarray}

At each receiving array element, echoes from all the transmitting elements are demodulated and then match-filtered respectively by their own carrier frequencies. For the $q$th sub-band, this procedure can be expressed by
\begin{eqnarray}
\label{Eq:DemodulationMatchedFiltering}
b_{p,l,r,q}(t; D,v,\alpha)&=& r_{p,l,r,q}(t;D,v,\alpha)\nonumber\\
&&\cdot\exp\big(-j2\pi(f_c+\Delta{f}[\mathbf{G}]_{p,l})t\big)\nonumber\\
&&\cdot\beta_{p,l,q}^{*}\exp(-j2\pi q\Delta{f}t).
\end{eqnarray}

Substituting (\ref{Eq:ReceivedSignal}) into (\ref{Eq:DemodulationMatchedFiltering}) gives the match-filtered sub-band echo:
\begin{eqnarray}
\label{Eq:BasebandSample}
b_{p,l,r,q}(D,v,\alpha)&=&\|\beta_{p,l,q}\|_2^2\nonumber\\
&&\cdot\exp\big(-j\frac{2\pi}{c}(f_c+\Delta{f}[\mathbf{G}_Q]_{pQ+q,l})\nonumber\\
&&\cdot(2D+2vpT+\alpha ld_T+\alpha rd_R)\big),
\end{eqnarray}
which is time-invariant. In one pulse train, the number of baseband samples (termed the measurement dimension) is $PQLR$. All the $PQLR$ samples can be arranged into a $(PQ)\times(LR)$ data matrix, whose $(pQ+q)$th row and $(lR+r)$th column entry is $b_{p,l,r,q}(D,v,\alpha)$. This data arrangement format is illustrated in Fig. \ref{Fig:DataMTX}.

As shown in (\ref{Eq:BasebandSample}), the baseband echoes of an RFD-MIMO radar depend simultaneously on the scatterer's range, velocity, and direction. Thus by vectorizing the data matrix, $\mathbf{u}(D,v,\alpha)\in\mathbb{C}^{(PQLR)\times 1}$ can be denoted as the \textit{range-velocity-direction} steering vector, where
\begin{equation}
\label{Eq:EleSV}
[\mathbf{u}(D,v,\alpha)]_{(lR+r)PQ+pQ+q}=b_{p,l,r,q}(D,v,\alpha).
\end{equation}

Furthermore, (\ref{Eq:BasebandSample}) also shows that the range-velocity-direction steering vector $\mathbf{u}(D,v,\alpha)$ can be decomposed into the Hadamard products of the modulation vector $\bm{\beta}$, the \textit{range} steering vector $\mathbf{u}_{\mathrm{D}}(D)$, the \textit{velocity} steering vector $\mathbf{u}_{\mathrm{V}}(v)$, and the \textit{direction} steering vector $\mathbf{u}_{\mathrm{A}}(\alpha)$:
\begin{equation}
\label{Eq:SVHMP}
\mathbf{u}(D, v, \alpha)=\bm{\beta}\odot\mathbf{u}_{\mathrm{D}}(D)\odot\mathbf{u}_{\mathrm{V}}(v)\odot\mathbf{u}_{\mathrm{A}}(\alpha),
\end{equation}
where
\begin{equation}
\label{Eq:ModulationVector}
[\bm{\beta}]_{(lR+r)PQ+pQ+q}=\|\beta_{p,l,q}\|_2^2, \forall r=0,1,\dots, R-1,
\end{equation}
\begin{equation}
\label{Eq:RangeSteeringVector}
\mathbf{u}_{\mathrm{D}}(D)=\vectorize\Big\{\exp\big(-j\frac{4\pi}{c}D(f_c+\Delta{f}\mathbf{G}_Q)\otimes \mathbf{1}^{1\times R}\big)\Big\},
\end{equation}
\begin{eqnarray}
\label{Eq:VelocitySteeringVector}
\mathbf{u}_{\mathrm{V}}(v)&=&\vectorize\Big\{\exp\big(-j\frac{4\pi}{c}Tv\big(f_c+\Delta{f}\mathbf{G}_Q\big)\nonumber\\
&&\odot(\mathbf{p}\otimes\mathbf{1}^{Q\times L})\otimes(\mathbf{1}^{1\times R})\big)\Big\},
\end{eqnarray}
and
\begin{eqnarray}
\label{Eq:DirectionSteeringVector}
\mathbf{u}_{\mathrm{A}}(\alpha)&=&\vectorize\Big\{\exp\big(-j\frac{2\pi}{c}\alpha(f_c+\Delta{f}\mathbf{G}_Q\otimes\mathbf{1}^{1\times R})\nonumber\\
&&\odot\big((d_T\mathbf{1}^{PQ\times 1}\otimes \mathbf{l}^T)\oplus (d_R\mathbf{r}^T)\big)\Big\}.
\end{eqnarray}
In the above equations, $\mathbf{p}=[0,1\dots, P-1]^T$, $\mathbf{l}=[0,1,\dots, L-1]^T$, and $\mathbf{r}=[0,1,\dots, R-1]^T$.

\begin{figure}[!h]
\centering
\includegraphics[width=.3\textwidth]{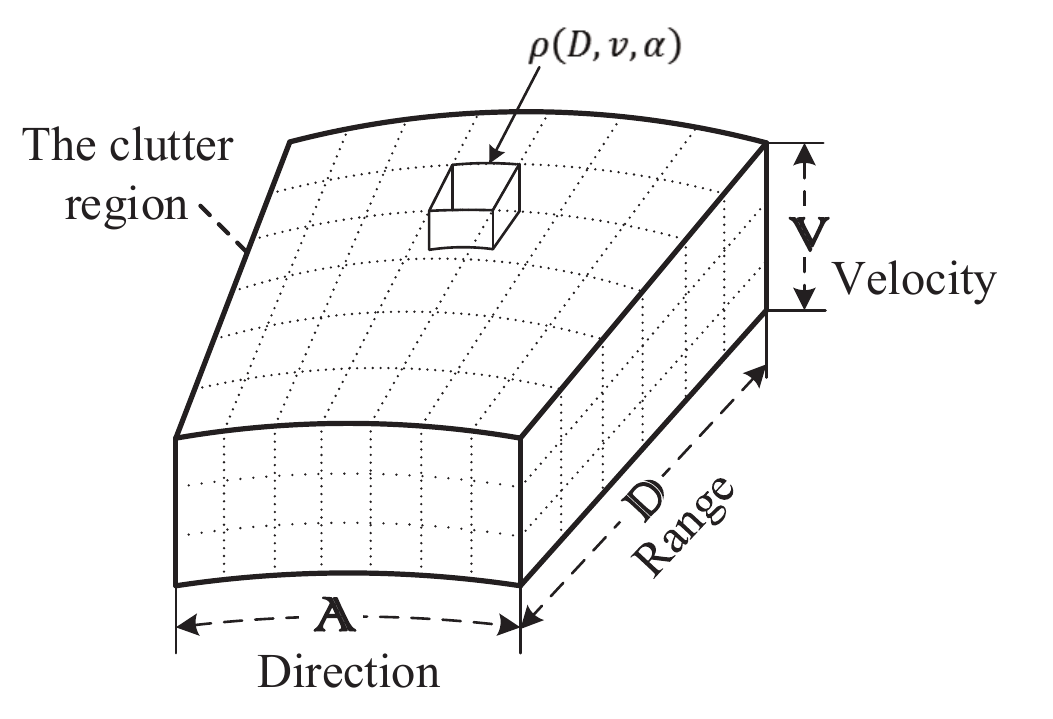}
\caption{A simple illustration of the clutter range-velocity-direction region, and the clutter reflection density.}
\label{Fig:ClutterRegion}
\end{figure}

In moving target indication (MTI)  \cite{Skolnik2002Introduction}, clutter, which is defined as the unwanted echo, is usually regarded as a superimposition of received echoes from scatterers whose ranges, velocities, and directions are in a certain region (the \textit{clutter region}). In this work, we denote $\mathbb{D}$, $\mathbb{V}$, and $\mathbb{A}$ as the clutter range, velocity, and direction regions, respectively. Hence the clutter echo vector can be calculated by
\begin{equation}
\label{Eq:ClutterEcho}
\mathbf{r}_{\mathrm{C}}=\int_\mathbb{D}\int_\mathbb{V}\int_\mathbb{A}\rho(D,v,\alpha)\cdot\mathbf{u}(D,v,\alpha)dDdvd\alpha,
\end{equation}
where $\rho(D,v,\alpha)$ is the clutter reflection density of the range-velocity-direction coordinate $\{D, v, \alpha\}$, as shown in Fig. \ref{Fig:ClutterRegion}.

The three-fold integral interval in ($\ref{Eq:ClutterEcho}$) is determined as follows.
\begin{enumerate}
\item
Clutter range region. Due to the clutter distribution features in practice \cite{Skolnik2002Introduction} and the narrowband assumption of each sub-band, clutter is usually combined with echoes from scatterers located over a large range. However, as can be seen in (\ref{Eq:RangeSteeringVector}), the values of the range steering vectors are identical\footnote{The phase factor $-j4\pi f_c D/c$ can be regarded as a part of the scatterer's reflection amplitude, because it remains constant w.r.t. different $p$, $q$, $r$, and $l$.}  for scatterers with range differences that are multiples of ${c}/(2\Delta{f})$. Thus the clutter range region is
\begin{equation}
\mathbb{D}=\big[0,\frac{c}{2\Delta{f}}\big].\nonumber
\end{equation}
\item
Clutter velocity region. Because the unambiguous target velocity is inversely proportional to the PRI in pulsed radars \cite{Skolnik2002Introduction}, the clutter velocity region should be a subset of the unambiguous interval of velocity:
\begin{eqnarray}
\mathbb{V}&\subseteq&\big[-\frac{c}{4f_cT},\frac{c}{4f_cT}\big].\nonumber
\end{eqnarray}
\item
Clutter direction region. As introduced in \cite{li2009mimo}, for a collocated MIMO radar, the clutter direction region should be a subset of the unambiguous interval of direction: 
\begin{eqnarray}
\mathbb{A}&\subseteq&[-\frac{c}{2d_Rf_c}, \frac{c}{2d_Rf_c}].\nonumber
\end{eqnarray}
\end{enumerate}

Moreover, the integral in (\ref{Eq:ClutterEcho}) can be approximated in a discrete mode by summing up the echoes from all the voxels in the clutter region. Assuming there are  $N_{\mathrm{C}}$ voxels, each of which has a reflection amplitude $\rho_n$ ($n=1,2\dots, N_{\mathrm{C}}$), then the clutter echo vector can be re-written as
\begin{equation}
\label{Eq:ClutterEchoMTX}
\mathbf{r}_{\mathrm{C}}=\mathbf{C}\cdot\bm{\rho},
\end{equation}
where $\bm{\rho}=[\rho_1,\rho_2,\dots, \rho_{N_{\mathrm{C}}}]^T$, and $\mathbf{C}$ is the clutter steering matrix (CSM), whose $n$th column is the steering vector of the range-velocity-direction coordinate $\{D_n,v_n,\alpha_n\}$.

\subsection{Application to Specific Frequency Diverse Waveforms}
\label{subsec:SignalModel:Substantiations}
The system and signal model presented in last subsection can be applied to specific FD waveforms by adapting corresponding parts of the model, such as $P$, $Q$, $L$, $R$, $d_R$, and $\mathbf{G}_Q$. Applications to FDA, SF,  and frequency diverse MIMO (FD-MIMO) radar waveforms will be derived in this subsection.

\subsubsection{FDA}
\label{subsubsec:FDA}
In radars using FDA antennas, different array elements transmit and receive different carrier frequencies which can be shifted linearly \cite{Antonik2006Frequency} (linear FDA, LFDA ) or randomly \cite{yimin2016ICASSP} (random FDA, RFDA). The range-direction dependent beampatterns are synthesized by processing the received echo.

In the FDA, the carrier frequencies of array elements are kept invariant throughout the operation. Thus the RFD-MIMO can be applied by the following steps.

First, reduce the FDCM, $\mathbf{G}$, to a row vector of $L$ entries, $\mathbf{g}^T$. Second, if the signal of a single pulse is wideband, the a-FDCM will be expressed by
\begin{equation}
\label{Eq:aFDCM_FDA}
\mathbf{G}_Q=\mathbf{q}\oplus\mathbf{g}^T.
\end{equation}

Because most research works about the FDA has focused on the beampatterns, the system models are usually formulated with one pulse, which makes the target velocity unobservable. Thus in (\ref{Eq:VelocitySteeringVector}), the variable $P$ should set to $1$, and the velocity steering vector $\mathbf{u}_{\mathrm{V}}(v)$ should be trivialized as an all-one vector. 

Moreover, each array element in an FDA transmits and receives with its own carrier frequency, thus in (\ref{Eq:TimeDelay}), $l=r$ and $d_T=d_R$. In addition, the clutter direction region should be adapted to $\mathbb{A}\subseteq[-c/(4d_Rf_c), c/(4d_Rf_c)]$. Hence the direction steering vector can be expressed by
\begin{eqnarray}
\label{Eq:DirectionSteeringVector}
\mathbf{u}_{\mathrm{A}}(\alpha)&=&\vectorize\Big\{\exp\big(-j\frac{4\pi}{c}\alpha\big(f_c+\Delta{f}(\mathbf{q}\oplus\mathbf{g}^T)\big)\nonumber\\
&&\odot(d_T\mathbf{1}^{Q\times 1}\otimes \mathbf{l}^T)\big)\Big\}.
\end{eqnarray}

Finally, the integral in (\ref{Eq:ClutterEcho}) should be calculated on only the clutter range region and clutter direction region. Therefore, the clutter model of an FDA radar is
\begin{equation}
\mathbf{r}_{\mathrm{C}}=\int_{\mathbb{D}}\int_{\mathbb{A}}\rho(D,\alpha)\cdot\bm{\beta}\odot\mathbf{u}_{\mathrm{D}}(D)\odot\mathbf{u}_{\mathrm{A}}(\alpha)dDd\alpha.\nonumber
\end{equation}

\subsubsection{SF pulse train}
\label{subsubsec:SF}
The applications to LSF and RSF pulse trains are straightforward. In radars with LSF or RSF pulse trains, the antennas are usually configured as single-input-single-output (SISO). Thus the FDCM should be reduced to a column vector $\mathbf{g}$ with $P$ entries, which represents the carrier frequencies of all pulses in a pulse train. For pulses with bandwidth $Q\Delta{f}$, the a-FDCM is given by
\begin{equation}
\label{Eq:aFDCM_SF}
\mathbf{G}_Q=\mathbf{g}\oplus\mathbf{q}.
\end{equation}

The single element antenna makes the target's direction unobservable in this instance. Hence in the waveform model, $L=R=1$, $d_R=d_R=0$, and the direction steering vector $\mathbf{u}_{\mathrm{D}}(D)$ should be replaced by an all-one vector, $\mathbf{1}^{(PQ)\times 1}$. With the above steps, the steering vectors of LSF and RSF pulse trains are range-velocity dependent:
\begin{equation}
\label{Eq:SteeringVectorSF}
\mathbf{u}(D, v)=\bm{\beta}\odot\mathbf{u}_{\mathrm{D}}(D)\odot\mathbf{u}_{\mathrm{V}}(v),
\end{equation}
where
\begin{eqnarray}
\label{Eq:RangeSteeringVectorSF}
\mathbf{u}_{\mathrm{D}}(D)&=&\vectorize\Big\{\exp\big(-j\frac{4\pi}{c}D(f_c+\Delta{f}\mathbf{g}\otimes \mathbf{q})\big)\Big\},\\
\label{Eq:VelocitySteeringVectorSF}
\mathbf{u}_{\mathrm{V}}(v)&=&\vectorize\Big\{\exp\big(-j\frac{4\pi}{c}Tv(f_c+\Delta{f}\mathbf{g}\otimes\mathbf{q})\nonumber\\
&&\odot(\mathbf{p}\otimes\mathbf{1}^{Q\times 1})\big)\Big\}.
\end{eqnarray}

Because it differs from that of an FDA radar, the clutter echo vector should be calculated by integrals on the clutter range region and clutter velocity region:
\begin{equation}
\mathbf{r}_{\mathrm{C}}=\int_{\mathbb{D}}\int_{\mathbb{V}}\rho(D,v)\cdot\bm{\beta}\odot\mathbf{u}_{\mathrm{D}}(D)\odot\mathbf{u}_{\mathrm{V}}(v)dDdv.\nonumber
\end{equation}

\subsubsection{FD-MIMO and its space-time adaptive processing (STAP) applications}
\label{subsubsec:FD-MIMO}
The MIMO technique has been applied to the FDA waveform to improve the measurement dimension \cite{sammartino2013frequency}. The FD-MIMO waveform is quite similar to the original RFD-MIMO model. However, most researches on FD-MIMO is focused on the range-direction dependent beampattern. Therefore, the a-FDCM of an FD-MIMO waveform can be written as
\begin{equation}
\label{Eq:aFDCM_MIMO}
\mathbf{G}_Q=\mathbf{q}\oplus\mathbf{g}^T.
\end{equation}

As the steering vector, the application can be accomplished by changing the velocity steering vector into an all-one vector, and removing the variable $v$ from the parameters of the steering vector.

However, in ground moving target indication (MTI) applications \cite{klemm2002principles}, such as space-time adaptive processing for airborne radars \cite{xu2015range}, the pulse train is introduced in the waveform, and the ground clutter's spatial frequency is assumed to be linearly proportional to the temporal frequency \cite{klemm2002principles}. For the most commonly studied side-looking mode antennas, the relationship between the spatial and temporal frequencies is
\begin{equation}
\label{Eq:STAPVariableDependency}
v=\alpha\cdot v_p,
\end{equation}
where $v_p$ is the platform velocity. Thus the velocity steering vector can be embedded into the direction steering vector, and the clutter model of an FD-MIMO radar with STAP applications can be formulated by substituting (\ref{Eq:STAPVariableDependency}) into (\ref{Eq:ClutterEcho}):
\begin{equation}
\label{Eq:ClutterEchoSTAP}
\mathbf{r}_{\mathrm{C}}=\int_\mathbb{D}\int_\mathbb{A}\rho(D,0,\alpha)\cdot\bm{\beta}\odot\mathbf{u}_{\mathrm{D}}(D)\odot\mathbf{u}_{\mathrm{V}}(\alpha v_p)\odot\mathbf{u}_{\mathrm{A}}(\alpha)dDd\alpha,\nonumber
\end{equation}
where $\mathbb{A}$ is the direction region covered by the array element. 

\section{Clutter Rank Estimation}
\label{Sec:ClutterRank}
Clutter rank, defined as the rank of a radar's CCM, is an important parameter for the quantification of target detection performance in clutter environments \cite{klemm2002principles,goodman2007clutter}. A small clutter rank relative to the whole measurement dimension means that the radar has a greater ability to suppress the clutter \cite{goodman2007clutter}. In this section, we explore the futures of the RFD-MIMO's CCM and CSM, and then give a theorem for the clutter rank estimation of the new waveform.

\subsection{Features of the CCM and CSM}
\label{subsec:ClutterRank:Matrix}
According to the definition of the CCM, we have that
\begin{equation}
\label{Eq:CCM}
\mathbf{R}_{\mathrm{C}}=E\big\{\mathbf{r}_{\mathrm{C}}\mathbf{r}_{\mathrm{C}}^H\big\}=\mathbf{C}E\left\{\bm{\rho}\bm{\rho}^H\right\}\mathbf{C}^H.
\end{equation}
From the basic properties of matrices \cite{gentle2007matrix}, the clutter rank is given by
\begin{eqnarray}
\label{Eq:CCMExp}
\mathcal{C}&\triangleq&\mathcal{R}\{\mathbf{R}_{\mathrm{C}}\}\nonumber\\
&\leq&\min\Big\{\mathcal{R}\{\mathbf{C}\},\mathcal{R}\big\{E\{\bm{\rho}\bm{\rho}^H\}\big\}\Big\}
\nonumber\\
&\leq& \mathcal{R}\{\mathbf{C}\},
\end{eqnarray}
which means the clutter rank of a radar is no larger than the rank of its CSM. The equality in the third row of (\ref{Eq:CCMExp}) is valid if the covariance matrix of the clutter reflection amplitudes, $E\{\bm{\rho}\bm{\rho}^H\}$, is full rank.  Furthermore, the rank of the CSM is
\begin{eqnarray}
\mathcal{R}\{\mathbf{C}\}&=&\mathcal{R}\{\mathbf{C}\mathbf{C}^H\}\nonumber\\
&=&\mathcal{R}\big\{\sum_{n=1}^{N_{\mathrm{C}}}\mathbf{u}(D_n, v_n, \alpha_n)\mathbf{u}^H(D_n, v_n, \alpha_n)\big\}.\nonumber
\end{eqnarray}

Then, the clutter rank estimation is relaxed to the rank estimation of the Gramian matrix of the CSM's Hermitian. Moreover, the summation in the above equation can be calculated by integrals on the clutter range, velocity, and direction regions:
\begin{equation}
\label{Eq:TripleInt}
\mathbf{C}\mathbf{C}^H=\int_\mathbb{D}\int_\mathbb{V}\int_\mathbb{A}\mathbf{u}(D_n, v_n, \alpha_n)\mathbf{u}^H(D_n, v_n, \alpha_n)dDdvd\alpha.
\end{equation}

By substituting (\ref{Eq:SVHMP}) into (\ref{Eq:TripleInt}), $\mathbf{C}\mathbf{C}^H$ can be decomposed into the Hadamard products of four matrices:
\begin{eqnarray}
\label{Eq:ClutterCovarianceHadamard}
\mathbf{C}\mathbf{C}^H&=&(\bm{\beta}\bm{\beta}^H)\odot\underbrace{\int_{\mathbb{D}}\mathbf{u}_{\mathrm{D}}(D)\mathbf{u}_{\mathrm{D}}^H(D)dD}_{\mathbf{R}_{\mathrm{D}}}\nonumber\\
&&\odot\underbrace{\int_{\mathbb{V}}\mathbf{u}_{\mathrm{V}}(v)\mathbf{u}_{\mathrm{V}}^H(v)dv}_{\mathbf{R}_{\mathrm{V}}}\odot \underbrace{\int_\mathbb{A}\mathbf{u}_{\mathrm{A}}(\alpha)\mathbf{u}_{\mathrm{A}}^H(\alpha)d\alpha}_{\mathbf{R}_{\mathrm{A}}}\nonumber\\
&=&(\bm{\beta}\bm{\beta}^H)\odot\mathbf{R}_{\mathrm{D}}\odot\mathbf{R}_{\mathrm{V}}\odot\mathbf{R}_{\mathrm{A}},
\end{eqnarray}
where $\bm{\beta}\bm{\beta}^H$ is rank-1, and $\mathbf{R}_{\mathrm{D}}$, $\mathbf{R}_{\mathrm{V}}$, and $\mathbf{R}_{\mathrm{A}}$ are all $(PQLR)\times(PQLR)$ matrices. Moreover, with the following lemma, it can be shown that the second component of (\ref{Eq:ClutterCovarianceHadamard}), $\mathbf{R}_{\mathrm{D}}$, has good features which can simplify the clutter rank estimation.

\begin{lemma}
\label{LE:RD}
The $a$th, $b$th entry of $\mathbf{R}_{\mathrm{D}}$, $[\mathbf{R}_{\mathrm{D}}]_{a,b}$, is non-zero, if and only if
\begin{equation}
\label{Eq:LE1Cond}
[\mathbf{G}_Q]_{I_{\mathrm{r}}(a),I_{\mathrm{c}}(a)}=[\mathbf{G}_Q]_{I_{\mathrm{r}}(b),I_{\mathrm{c}}(b)},
\end{equation}
where 
\begin{eqnarray}
I_{\mathrm{r}}(x) &=& x-(PQ)\cdot\lfloor x/(PQ)\rfloor, \nonumber\\
I_{\mathrm{c}}(x) &=& \lfloor x/(PQ)\rfloor. \nonumber
\end{eqnarray}
\end{lemma}

\textbf{Proof:}  According to the definition of $\mathbf{R}_{\mathrm{D}}$ in (\ref{Eq:ClutterCovarianceHadamard}),
\begin{eqnarray}
[\mathbf{R}_{\mathrm{D}}]_{a,b}&=&\int_{\mathbb{D}}[\mathbf{u}_{\mathrm{D}}(D)]_a[\mathbf{u}_{\mathrm{D}}(D)]_b^{\ast}dD\nonumber\\
&=&\int_{\mathbb{D}}e^{-j\frac{4\pi}{c}Dz}dD,\nonumber
\end{eqnarray}
where $z\in\mathbb{Z}$ and $z =[\mathbf{G}_Q]_{I_r(a),I_c(a)}-[\mathbf{G}_Q]_{I_r(b),I_c(b)}$. If the condition given in (\ref{Eq:LE1Cond}) is satisfied,
\begin{eqnarray}
[\mathbf{R}_{\mathrm{D}}]_{a,b}=\int_{0}^{\frac{c}{2\Delta{f}}}e^{-j\frac{4\pi\Delta{f}}{c}D\cdot 0}dD
=\frac{c}{2\Delta{f}}.\nonumber
\end{eqnarray}
Otherwise, if 
\begin{equation}
[\mathbf{G}_Q]_{I_r(a),I_c(a)}\neq[\mathbf{G}_Q]_{I_r(b),I_c(b)},
\end{equation}
$z $ becomes a non-zero integer. According to the Cauchy's integral theorem \cite{rudin1987real},
\begin{equation}
[\mathbf{R}_{\mathrm{D}}]_{a,b}=\int_{0}^{\frac{c}{2\Delta{f}}}e^{-j\frac{4\pi\Delta{f}}{c}D\cdot z}dD=0.\nonumber
\end{equation}
Lemma \ref{LE:RD} is proven.\hfill{$\square$}

Lemma \ref{LE:RD} and (\ref{Eq:ClutterCovarianceHadamard}) mean that many entries of $\mathbf{C}\mathbf{C}^H$ are zero. Meanwhile, because $\mathbf{R}_{\mathrm{D}}$ is symmetric, the rest of the non-zero entries can be permuted into a block diagonal matrix by row and column swapping, where each diagonal block corresponds to a frequency point, $f_c+m\Delta{f}$ ($m\in\mathbb{M}_Q$). Moreover, because the rank of a matrix remains unchanged during the row and column swapping, $\mathcal{R}\{\mathbf{C}\mathbf{C}^H\}$ can be decomposed to the sum of several smaller matrices' ranks:
\begin{equation}
\label{Eq:BlockDiag}
\mathcal{R}\{\mathbf{C}\mathbf{C}^H\}=\sum_{m\in\mathbb{M}_Q}\mathcal{R}\big\{\mathbf{R}_{\mathrm{C}_m}\big\},
\end{equation}
where $\mathbf{R}_{\mathrm{C}_m}$ is the diagonal block corresponding to the frequency point $f_c+m\Delta{f}$.
\begin{figure}[!h]
\centering
\includegraphics[width=.5\textwidth]{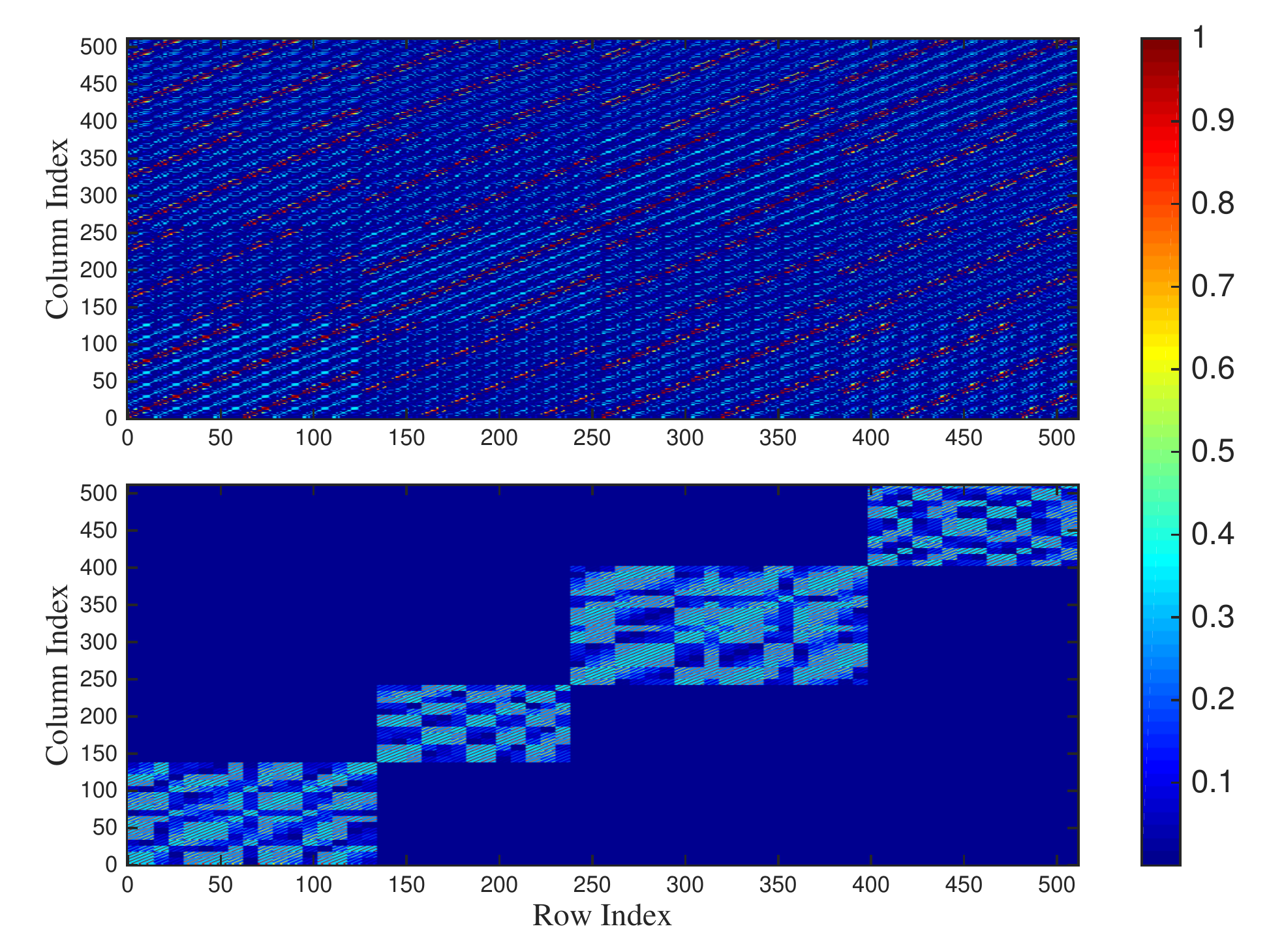}
\caption{An example of the entries' magnitudes in $\mathbf{C}\mathbf{C}^H$, before (upper half) and after (lower half) row and column swapping.}
\label{Fig:CCM}
\end{figure}

Fig. \ref{Fig:CCM} shows an example for the sparse and symmetric features of $\mathbf{C}\mathbf{C}^H$. In this example, the RFD-MIMO waveform had four different carrier frequencies and 16 monotone pulses in one CPI. The numbers of transmitting and receiving array elements were four and eight, respectively. The upper half of Fig. \ref{Fig:CCM} shows the entries' magnitudes in the original $\mathbf{C}\mathbf{C}^H$, and the lower half displays those of the permuted matrix. It can be clearly seen that after the row and column swapping, the permuted $\mathbf{C}\mathbf{C}^H$ has a block diagonal structure of four diagonal blocks. 

In accordance with the row and column swapping, each $\mathbf{R}_{\mathrm{C}_{m}}$ can be decomposed to a Hadamard product of two matrices:
\begin{equation}
\label{Eq:Hardamad}
\mathbf{R}_{\mathrm{C}_m}=\mathbf{R}_{\mathrm{V}_m}\odot\mathbf{R}_{\mathrm{A}_m},
\end{equation}
where
\begin{eqnarray}
\label{Eq:ClutterCovarianceSub}
\mathbf{R}_{\mathrm{V}_m}&=&\int_{\mathbb{V}}\mathbf{u}_{\mathrm{V}_m}(v)\mathbf{u}^H_{\mathrm{V}_m}(v)dv,\\
\label{Eq:ClutterCovarianceSub2}
\mathbf{R}_{\mathrm{A}_m}&=&\int_{\mathbb{A}}\mathbf{u}_{\mathrm{A}_m}(\alpha)\mathbf{u}^H_{\mathrm{A}_m}(\alpha)d\alpha.
\end{eqnarray}
In (\ref{Eq:ClutterCovarianceSub}-\ref{Eq:ClutterCovarianceSub2}), $\mathbf{u}_{\mathrm{V}_m}$ and $\mathbf{u}_{\mathrm{A}_m}$ are sub-velocity and sub-direction steering vectors, given by
\begin{eqnarray}
\label{Eq:BlockSteering}
\mathbf{u}_{\mathrm{V}_m}(v)&=&\exp\Big(-j\frac{4\pi}{c}v(f_c+m\Delta{f})\nonumber\\
&&\cdot\big((T\lfloor\frac{\mathcal{I}^{\parallel}_m(\mathbf{G}_Q)}{Q}\rfloor)\otimes \mathbf{1}^{R\times 1}\big)\Big),\\
\label{Eq:BlockSteering2}
\mathbf{u}_{\mathrm{A}_m}(\alpha)&=&\exp\Big(-j\frac{2\pi}{c}\alpha(f_c+m\Delta{f})\nonumber\\
&&\cdot\big((d_T\mathcal{I}^{=}_m(\mathbf{G}_Q))\oplus (d_R\mathbf{r})\big)\Big).
\end{eqnarray}

With above derivations, $\mathbf{C}\mathbf{C}^H$ can be expressed by small matrices, such as $\mathbf{R}_{\mathrm{V}_m}$ and $\mathbf{R}_{\mathrm{A}_m}$, whose dimensions are notably reduced from the original ones. In addition, each $\mathbf{R}_{\mathrm{V}_m}$ and $\mathbf{R}_{\mathrm{A}_m}$ depends only on a single frequency point, $f_c+m\Delta{f}$. These features allow the clutter rank estimation of a diverse waveform to be accomplished in a ``frequency point by frequency point" manner, which greatly reduces the complexity of the original problem.

\subsection{Rank Estimation}
\label{subsec:ClutterRank:Rank}
The integrals in (\ref{Eq:ClutterCovarianceSub}) and (\ref{Eq:ClutterCovarianceSub2}) can be approximated in matrix form, by discretizing the clutter velocity region and clutter direction region into $N_{\mathrm{V}}$ velocity grids and $N_{\mathrm{A}}$ direction grids, respectively:
\begin{eqnarray}
\mathbf{R}_{\mathrm{V}_m}&=&\mathbf{C}_{\mathrm{V}_m}\mathbf{C}_{\mathrm{V}_m}^H,\nonumber\\
\mathbf{R}_{\mathrm{A}_m}&=&\mathbf{C}_{\mathrm{A}_m}\mathbf{C}_{\mathrm{A}_m}^H,\nonumber
\end{eqnarray}
where
\begin{equation}
\label{Eq:SteerSampleVelocity}
[\mathbf{C}_{\mathrm{V}_m}]_n=\mathbf{u}_{\mathrm{V}_m}(v_n), \mathrm{ } n=1,2,\dots,N_{\mathrm{V}},
\end{equation}
\begin{equation}
\label{Eq:SteerSampleDirection}
[\mathbf{C}_{\mathrm{A}_m}]_n=\mathbf{u}_{\mathrm{A}_m}(\alpha_n), \mathrm{ } n=1,2,\dots,N_{\mathrm{A}}.
\end{equation}

Take the rank estimation of $\mathbf{R}_{\mathrm{V}_m}$ as an example. According to the basic properties of matrices \cite{gentle2007matrix}, $\mathcal{R}\{\mathbf{R}_{\mathrm{V}_m}\}=\mathcal{R}\{\mathbf{C}_{\mathrm{V}_m}\}$. Furthermore, (\ref{Eq:BlockSteering}) and (\ref{Eq:SteerSampleVelocity}) show that the columns of $\mathbf{C}_{\mathrm{V}_m}$ can be regarded as complex sinusoids sampled on the $m$th \textit{temporal sampling aperture}, which is defined as the set composed of all the unique entries in vector $T\lfloor{\mathcal{I}^{\parallel}_m(\mathbf{G}_Q)}/{Q}\rfloor$, in increasing order.

Some characteristics of the sampling apertures should be noted. First, if the a-FDCM has multiple non-identical columns, the sampling apertures corresponding to different frequency points may overlap. Second, each sampling aperture can be divided into sub-apertures by splitting two successive sampling instants into two sub-apertures when the gap between these two instants is larger than the Nyquist sampling interval, ${c}/({2(f_c+m\Delta{f})\langle\mathbb{V}\rangle})$. All the sub-apertures corresponding to the frequency point, $f_c+m\Delta{f}$, are gathered as a set, ${\mathbb{T}}_m({\mathbb{V}})$.

From the above discussion, an approximation of $\mathcal{R}\{\mathbf{R}_{\mathrm{V}_m}\}$ can be provided with the help of prolate spheroidal wave function (PSWF) theory \cite{landau1962prolate}. According to PSWF theory, complex sinusoids, whose energies are mostly confined in a certain ``time($T$)-frequency($W$)" region, can be well approximated by linear combinations of $\lceil WT + 1\rceil$ orthogonal functions. In our case, the ``frequencies" of the complex sinusoids, $\mathbf{u}_{\mathrm{V}_m}(v_n)$, vary in an extent of $2(f_c+m\Delta{f})\langle\mathbb{V}\rangle/c$. The ``time" should be counted separately for every sub-aperture in the ${\mathbb{T}}_m({\mathbb{V}})$. Then we have the following lemma.
\begin{lemma}
\label{LE:RankEvLRV}
The rank of $\mathbf{R}_{\mathrm{V}_m}$ can be approximated by
\begin{equation}
\label{Eq:RankV}
\mathcal{R}\{\mathbf{R}_{\mathrm{V}_m}\}\approx\mathcal{U}_{\mathrm{V}_m}\triangleq\min\big\{|\mathcal{I}_m^{\parallel}\{\mathbf{G}_Q\}|,R_{\mathrm{V}_m}(\mathbb{V}),\tilde{R}_{\mathrm{V}_m}(\mathbb{V})\big\},
\end{equation}
where
\begin{equation}
R_{\mathrm{V}_m}(\mathbb{V})=\lceil\frac{2}{c}(f_c+m\Delta{f})\langle\mathbb{V}\rangle\langle T\lfloor \frac {\mathcal{I}_m^{\parallel}\{\mathbf{G}_Q\}}{Q}\rfloor\rangle\rceil+1,
\end{equation}
and
\begin{equation}
\tilde{R}_{\mathrm{V}_m}(\mathbb{V}) =\sum_{{\mathcal{T}}\in{\mathbb{T}}_m({\mathbb{V}})}\lceil\frac{2}{c}(f_c+m\Delta{f})\langle\mathbb{V}\rangle\langle{\mathcal{T}}\rangle\rceil+1.
\end{equation}
\end{lemma}
In equation (\ref{Eq:RankV}), the first and second terms in the function $\min\{\cdot\}$  are needed because the maximal number of linearly dependent signals with identical sampling instants is no larger than the number of the sampling instants, and cannot be reduced by introducing new sampling instants. 
\begin{figure}[!t]
\centering
\includegraphics[width=.5\textwidth]{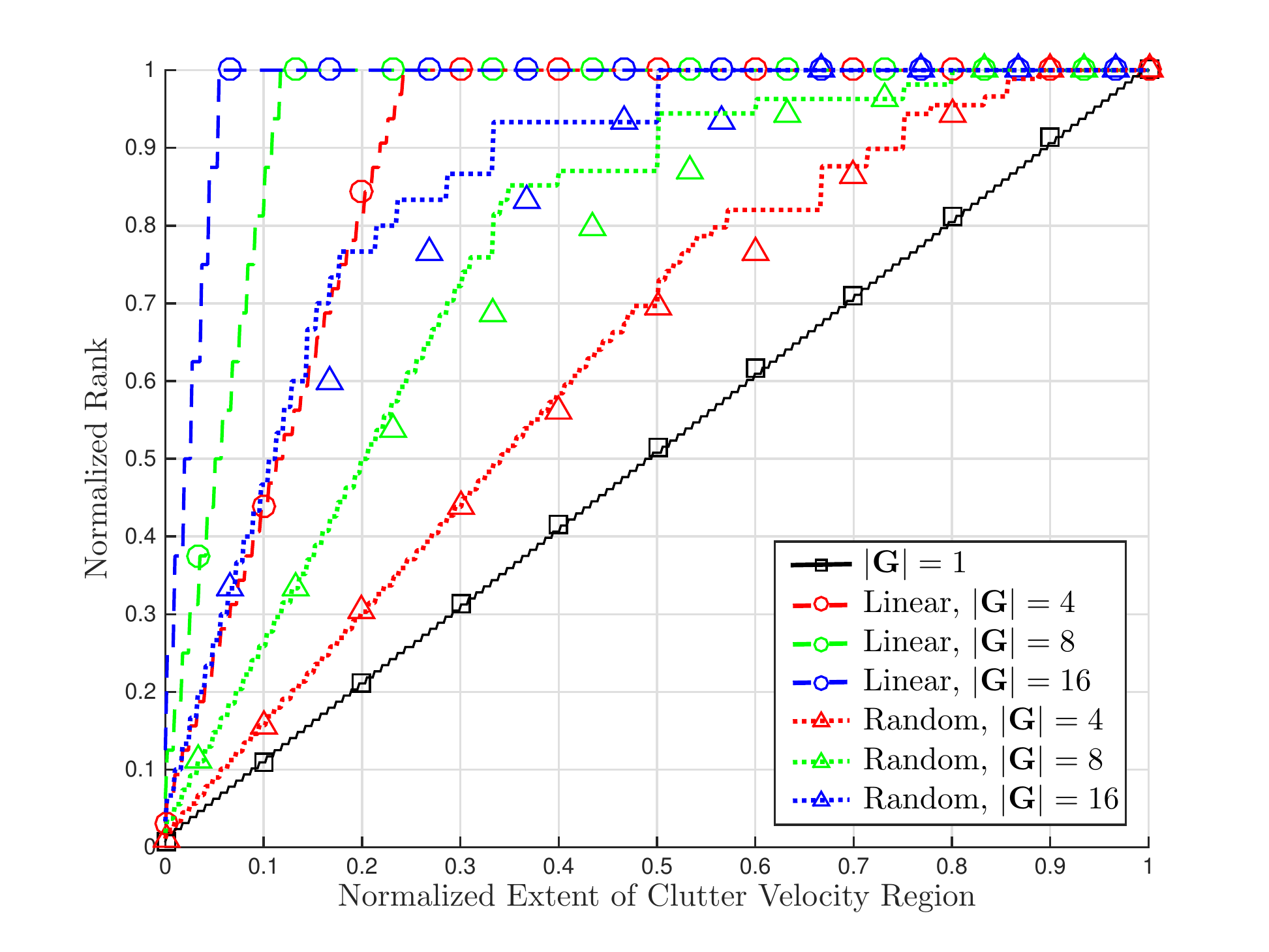}
\caption{The normalized $\mathcal{U}_{\mathrm{V}_m}$ and the ranks of $\mathbf{R}_{\mathrm{V}_m}$, for different normalized extents of clutter velocity region.}
\label{Fig:Rank_Rm_V}
\end{figure}

An example of $\mathcal{R}\{\mathbf{R}_{\mathrm{V}_m}\}$ and $\mathcal{U}_{\mathrm{V}_m}$ is provided in Fig. \ref{Fig:Rank_Rm_V}. In this example, we formulated an RFD-MIMO waveform with 32 transmitting array elements, 8 receiving array elements, and 128 monotone pulses. Both linear and random carrier frequency assignments were considered. The numbers of carrier frequencies were set as $|\mathbf{G}|=1$ (the fixed-frequency case, expressed similarly hereinafter), $4$, $8$, and $16$. In this simulation, the extent of the clutter velocity region varied from zero to one times the unambiguous extent of velocity. Both $\mathcal{R}\{\mathbf{R}_{\mathrm{A}_m}\}$ and $\mathcal{U}_{\mathrm{V}_m}$ were normalized by the numbers of unique sampling instants in the temporal sampling aperture, and are indicated by symbols and dotted/dashed lines, respectively. The different colors indicate to different carrier frequency numbers. The circles and triangles represent linear and random carrier frequency assignments, respectively. In this simulation, the approximations given by Lemma \ref{LE:RankEvLRV} matched the true ranks well. In addition, other phenomena could be seen: The fixed-frequency waveforms had the smallest rank; the larger the carrier frequency number, the faster the normalized rank grew; and the ranks of random carrier frequency assignments were smaller than those of the linear ones.

The rank estimation of $\mathbf{R}_{\mathrm{A}_m}$ can be derived in a similar way. As shown in (\ref{Eq:BlockSteering2}) and (\ref{Eq:SteerSampleDirection}), the columns of $\mathbf{C}_{\mathrm{A}_m}$ can be regarded as the discrete samples of complex sinusoids whose ``frequencies" vary in an extent of $(f_c+m\Delta{f})\langle\mathbb{A}\rangle/c$. Moreover, the sampling instants are distributed on the $m$th \textit{spatial sampling aperture}, which is determined by $\langle(d_T\mathcal{I}_m^{=}\{\mathbf{G}_Q\})\oplus(d_R\mathbf{r})\rangle$. The Nyquist sampling interval in the spatial domain is ${c}/({(f_c+m\Delta{f})\langle\mathbb{A}d_R\rangle})$. Thus the sampling instants corresponding to $f_c+m\Delta{f}$ can be divided to sub-apertures, all of which are gathered as a  set, ${\mathbb{S}}_m({\mathbb{A}})$. With the above definitions, we have the following lemma.
\begin{lemma}
\label{LE:RankSupRA}
The rank of $\mathbf{R}_{\mathrm{A}_m}$  can be approximated by
\begin{eqnarray}
\label{Eq:RankA}
\mathcal{R}\{\mathbf{R}_{\mathrm{A}_m}\}&\approx&\mathcal{U}_{\mathrm{A}_m}\nonumber\\
&\triangleq&\min\big\{|(d_T\mathcal{I}_m^{=}\{\mathbf{G}_Q\})\oplus(d_R\mathbf{r})|, R_{\mathrm{A}_m}(\mathbb{A}),\nonumber\\
&&\tilde{R}_{\mathrm{A}_m}(\mathbb{A})\big\},
\end{eqnarray}
where
\begin{equation}
R_{\mathrm{A}_m}(\mathbb{A}) = \lceil\frac{1}{c}(f_c+m\Delta{f})\langle\mathbb{A}\rangle\langle( d_T\mathcal{I}_m^{=}\{\mathbf{G}_Q\})\oplus(d_R\mathbf{r})\rangle\rceil+1,
\end{equation}
and
\begin{equation}
\tilde{R}_{\mathrm{A}_m}(\mathbb{A}) =\sum_{\mathcal{S}\in{\mathbb{S}}_m({\mathbb{A}})}\lceil\frac{1}{c}(f_c+m\Delta{f})\langle\mathbb{A}\rangle\langle{\mathcal{S}}\rangle\rceil+1.
\end{equation}
\end{lemma}

\begin{figure}[!t]
\centering
\includegraphics[width=.5\textwidth]{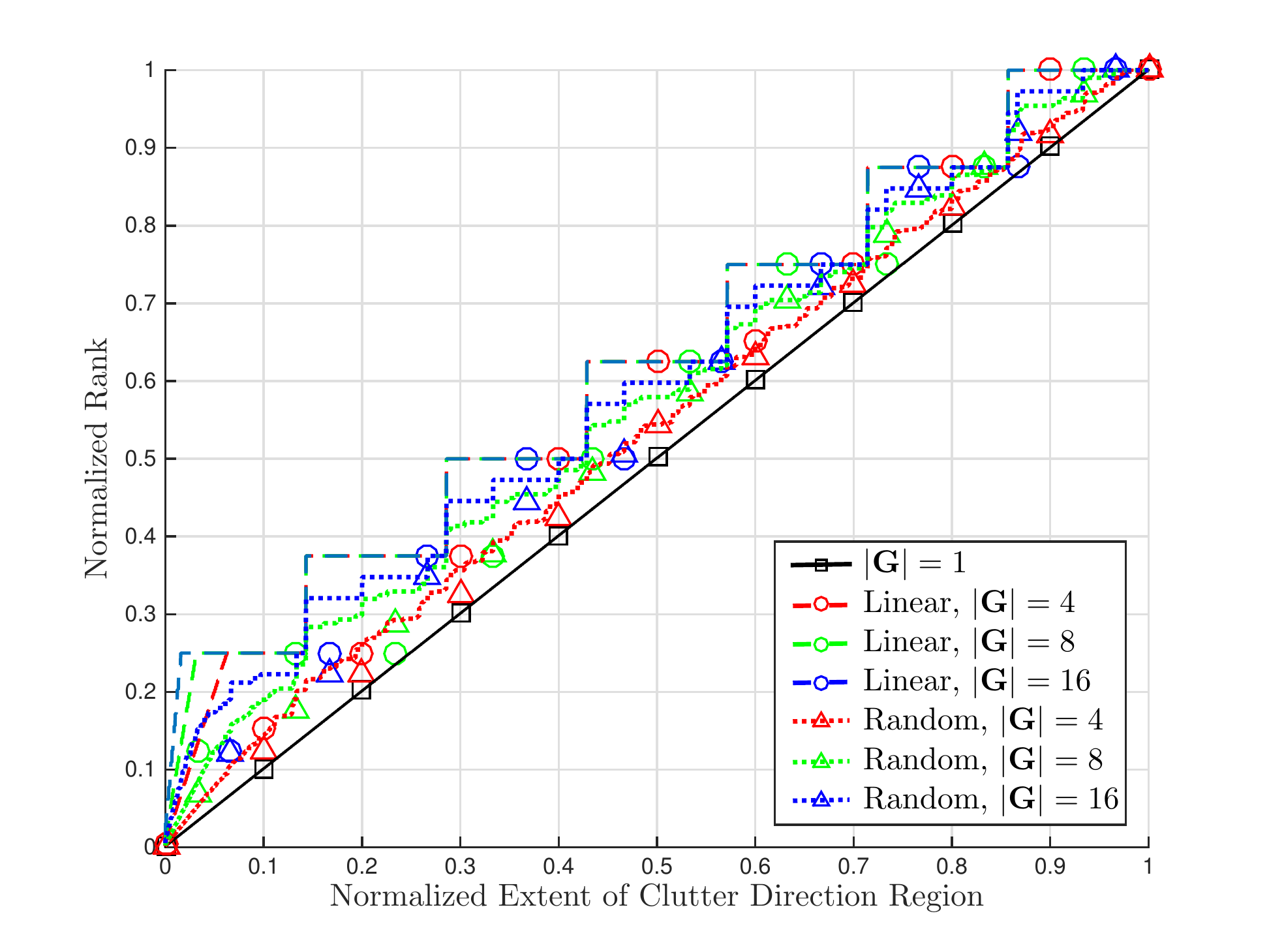}
\caption{The normalized $\mathcal{U}_{\mathrm{A}_m}$ and the normalized ranks of $\mathbf{R}_{\mathrm{A}_m}$, for different normalized extents of clutter direction region.}
\label{Fig:Rank_Rm_A}
\end{figure}

An example of $\mathcal{R}\{\mathbf{R}_{\mathrm{A}_m}\}$ and $\mathcal{U}_{\mathrm{A}_m}$ is given in Fig. \ref{Fig:Rank_Rm_A}. The simulation setup and legends are the same as those in Fig. \ref{Fig:Rank_Rm_V}, except the pulse number is 32 and the transmitting elements number is 128. Although the system configurations were similar to those of the simulation for $\mathcal{R}\{\mathbf{R}_{\mathrm{V}_m}\}$, the results were different. In this example, the clutter ranks corresponding to different carrier frequency numbers had only small differences between each other, for both linear and random carrier frequency assignments.

The reason can be briefly explained as follows: In the MIMO antenna, the spatial sampling aperture of each frequency point is the stretched sum of the transmitting aperture  ($d_T\mathcal{I}_m^{=}\{\mathbf{G}_Q\}$) and the receiving aperture ($d_R\mathbf{r}$). Because $d_T$ is usually several times larger than $d_R$, the discontinuity of entries in $\mathcal{I}_m^{=}\{\mathbf{G}_Q\}$ leads to a wide gap between spatial sampling instants. Hence when $\langle\mathbb{A}\rangle$ is small, the spatial sampling aperture begins to divide into small sub-apertures. These sub-apertures are composed of integer multiples of $R$ sampling instants, and the distances between neighboring sampling instants are $d_R$. Therefore, the rank is approximately proportional to the number of sampling instants, regardless of the number and the assignment of  carrier frequencies.

With Lemma \ref{LE:RankEvLRV} and Lemma \ref{LE:RankSupRA}, we have the following lemma for the rank of $\mathbf{R}_{\mathrm{C}_m}$.
\begin{lemma}
\label{LE:RankSupRC}
The ranks of the diagonal blocks, $\mathbf{R}_{\mathrm{C}_m}$, satisfy the following inequalities:
\begin{equation}
\label{Eq:BlockRank}
\mathcal{U}_{\mathrm{C}_m}\leq\mathcal{R}\{\mathbf{R}_{\mathrm{C}_m}\}\leq\min\{K_m, \mathcal{U}_{\mathrm{V}_m}\mathcal{U}_{\mathrm{A}_m}\},
\end{equation}
where
\begin{equation}
\label{Eq:DefineUC}
\mathcal{U}_{\mathrm{C}_m}\triangleq\min\big\{K_m,\mathcal{U}_{\mathrm{V}_m}+\mathcal{U}_{\mathrm{A}_m}-1\big\},
\end{equation}
and $K_m$ is the dimension of $\mathbf{R}_{\mathrm{C}_m}$.
\end{lemma}

\textbf{Proof:} The proof can be found in Appendix \ref{AP:One}. \hfill{$\square$}

With the above preparation, we deduce the following theorem for the clutter rank estimation of the RFD-MIMO waveform:
\begin{theorem}
\label{TH:ClutterRank}
If the covariance matrix of the clutter reflection amplitudes is full rank, then the clutter rank of an RFD-MIMO radar, $\mathcal{C}$, satisfies the following inequalities:
\begin{eqnarray}
\label{Eq:Th}
\sum_{m\in\mathbb{M}_Q}\mathcal{U}_{\mathrm{C}_m}\leq\mathcal{C}\leq\sum_{m\in\mathbb{M}_Q}\min\{K_m, \mathcal{U}_{\mathrm{V}_m}\mathcal{U}_{\mathrm{A}_m}\},
\end{eqnarray}
where 
$\mathcal{U}_{\mathrm{V}_m}$, $\mathcal{U}_{\mathrm{A}_m}$, and $\mathcal{U}_{\mathrm{C}_m}$ are defined in (\ref{Eq:RankV}), (\ref{Eq:RankA}), and (\ref{Eq:DefineUC}), respectively.
\end{theorem}

\textbf{Proof:} The proof of Theorem \ref{TH:ClutterRank} can be accomplished straightforwardly by combining (\ref{Eq:BlockDiag}) and Lemma \ref{LE:RankSupRC}. \hfill{$\square$}

\section{Applications and Discussion}
\label{Sec:Extension}
Due to the flexible configuration of the a-FDCM, the RFD-MIMO waveform is a general model for FD waveforms whose carrier frequencies can vary for different array elements and/or pulses. In this section, we will give the rank estimation for specific kinds of FD waveforms by the corollaries of Theorem \ref{TH:ClutterRank}.

\subsection{Metrics of Clutter Suppression Performance}
\label{subsec:ClutterRank:performance}
As concluded in \cite{goodman2007clutter}, a higher clutter rank relative to the measurement dimension means that the clutter spreads over a larger portion of the whole signal space, leaving fewer clutter-free dimensions for the target detection. In this section, we will compare the normalized clutter rank (NCR, defined as the clutter rank, $\mathcal{C}$, normalized by the measurement dimension, $PQLR$) between different FD radar waveforms, to show the corresponding clutter suppression potentials. The reasons for choosing the NCR are as follows.

First, as explained in Appendix \ref{AP:Two}, the $\mathrm{SCNR}_{\mathrm{opt}}$, defined as the optimal output signal-to-clutter-noise-ratio which can be achieved by linear filtering, is approximately proportional to the target energy which is spread in the orthogonal complement of the CCM's eigenspace:
\begin{equation}
\label{Eq:OptSCNRVec}
\mathrm{SCNR}_{\mathrm{opt}}\approx\frac{1}{\sigma^2}\|\mathbf{P}_{\mathbf{R}_{\mathrm{C}}}^{\perp}\cdot\mathbf{u}(D,v,\alpha)\|_2^2,
\end{equation}
where $\mathbf{P}_{\mathbf{R}_{\mathrm{C}}}^{\perp}$ is the projection matrix.

Numerical investigations (not analytically proven yet) showed that the averaged (w.r.t. all the unambiguous extents of target range, velocity, and direction) target power distributed on the CCM's  eigenspace is linearly proportional to the normalized clutter rank (NCR):
\begin{equation}
\label{Eq:ClutterProjRatio}
\mean_{D,v,\alpha}\big\{\|\mathbf{P}^{\perp}_{\mathbf{R}_{\mathrm{C}}}\cdot\mathbf{u}(D,v,\alpha)\|_2^2\big\}\varpropto1-\frac{\mathcal{C}}{PQLR}.
\end{equation}
The result in (\ref{Eq:ClutterProjRatio}) implies that the averaged $\mathrm{SCNR}_{\mathrm{opt}}$ can be predicted by the difference between the quantity $1$ and the NCR. 

Furthermore, the frequency diversity loss (FDL), defined as the ratio between the $\mathrm{SCNR}_{\mathrm{opt}}$ of an FD waveform and the $\mathrm{SCNR}_{\mathrm{opt}}$ of a fixed-frequency waveform, can be expressed by the NCR:
\begin{equation}
\label{Eq:DefDiverseLoss}
\mathcal{L}_{\mathrm{FD}}\approx\frac{1-\mathcal{C}_{\mathrm{FD}}/(PQLR)}{1-\mathcal{C}_0/(PQLR)},
\end{equation}
where $\mathcal{L}_{\mathrm{FD}}$ is the FDL, and $\mathcal{C}_{\mathrm{FD}}$ and $\mathcal{C}_0$ are the clutter ranks of the fixed-frequency and the FD waveforms for the same clutter environment\footnote{It should be noted that in this work, the clutter suppression is accomplished by coherent processing, thus the results defer from the traditional incoherent cases. In addition, clutter whose delay is larger than one CPI is unconsidered.}.  

\subsection{Frequency Diverse Array}
As introduced in subsection \ref{subsubsec:FDA}, the applications of the RFD-MIMO to FDA can be accomplished by adapting the matrix $\mathbf{G}$ to a row vector $\mathbf{g}^T$. Thus according to Theorem \ref{TH:ClutterRank},  the clutter rank of an FDA radar can be evaluated by the following corollary.
\begin{corollary}
\label{Co:FDA}
(\textit{Frequency Diverse Array}) For an FDA radar with a clutter direction region $\mathbb{A}$, the clutter rank is
\begin{eqnarray}
\label{Eq:RankFDA1}
\mathcal{C}&\approx&\sum_{m\in\mathbb{M}_Q}\min\Big\{|\mathcal{I}_m^{=}\{\mathbf{q}\oplus\mathbf{g}^T\}|, \nonumber\\
&&\lceil\frac{2}{c}(f_c+m\Delta{f})\langle d_T\mathcal{I}_m^{=}\{\mathbf{q}\oplus\mathbf{g}^T\}\rangle\cdot\langle\mathbb{A}\rangle\rceil+1,\nonumber\\
&&\sum_{\mathcal{S}\in\mathbb{S}_m(\mathbb{A})}\lceil\frac{2}{c}(f_c+m\Delta{f})\langle\mathbb{A}\rangle\langle\mathcal{S}\rangle\rceil+1\Big\}.
\end{eqnarray}
\end{corollary}

\begin{figure}[!t]
\centering
\includegraphics[width=.5\textwidth]{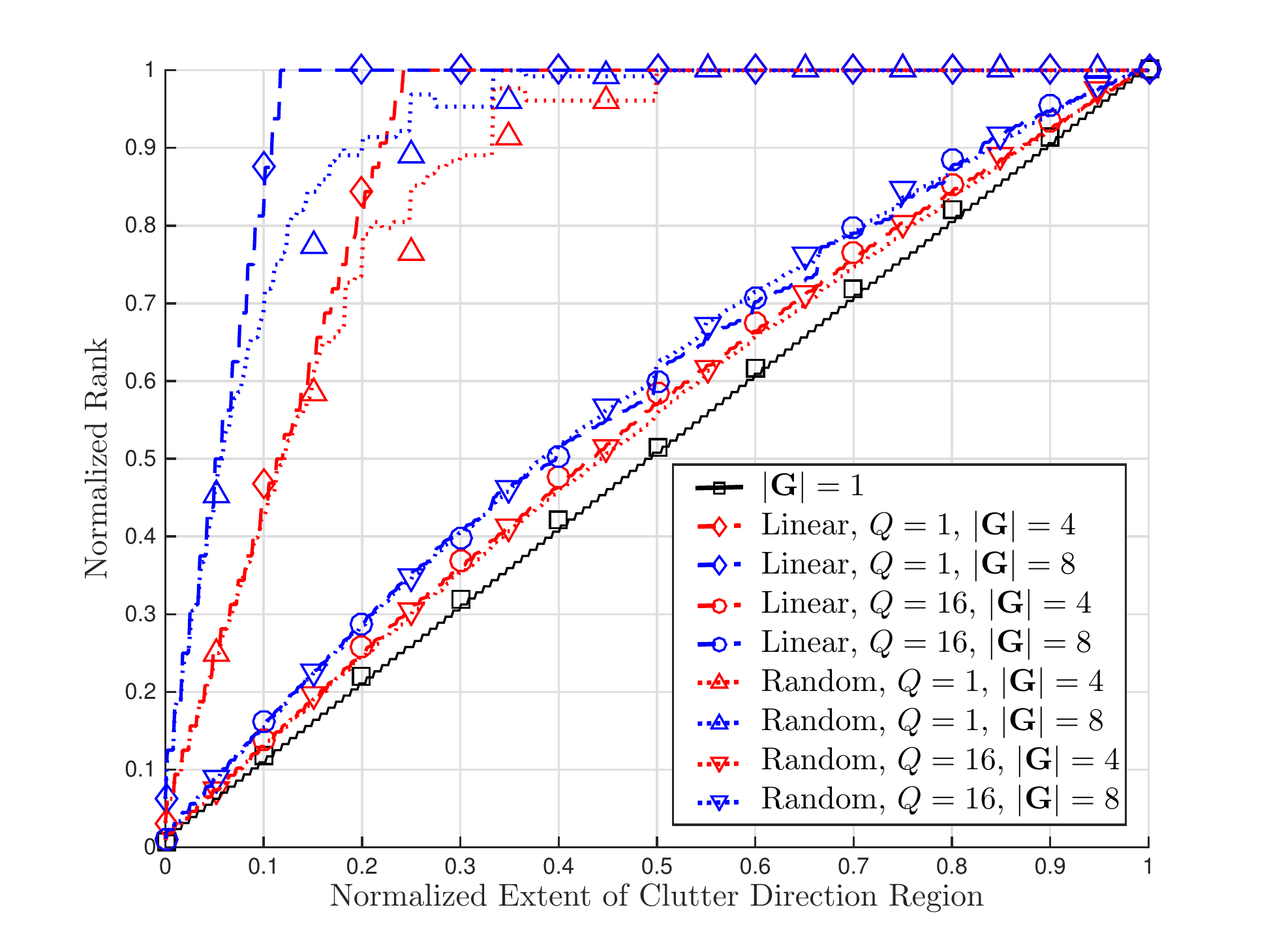}
\caption{NCRs and their approximations for the FDA radar waveforms.}
\label{Fig:Rank_All_FDA}
\end{figure}

The NCRs of both LFDA and RFDA are illustrated in Fig. \ref{Fig:Rank_All_FDA}. In this example, there were $L=256$ array elements, and the carrier frequency of each element was selected from a set of $|\mathbf{G}|=1$, $4$, and $8$ integers. Both monotone ($Q=1$) and wideband ($Q=16$) pulses were simulated. The clutter rank $\mathcal{C}$ and its approximation given in (\ref{Eq:RankFDA1}) were calculated and then normalized by the measurement dimension $PQLR$. In the figure, the NCRs for different system configurations are indicated by different symbols, and the approximations are plotted by dashed and dotted lines for the LFDA and the RFDA, respectively.

It is shown that the fixed-frequency ($|\mathbf{G}|=1$) waveforms have the lowest NCRs, and that the lower the carrier frequency number, the smaller the NCR. In addition, the NCRs of wideband pulse waveforms ($Q=16$) are much smaller than those of the monotone ($Q=1$) ones. Furthermore, the LFDA has a higher NCR than the RFDA, especially when $\langle\mathbb{A}\rangle$ is in the intermediate portion of the normalized extent of the clutter direction region. This phenomenon will be further discussed in subsection \ref{subsec:discussion}, together with the SF pulse trains.
\subsection{Stepped-Frequency Pulse Train}
Coherent clutter suppression and moving target indication are long-term problems for the SF, especially for the RSF (also known as frequency agile coherent \cite{ghelfi2012phase}) radars \cite{Skolnik2002Introduction}. With Theorem \ref{TH:ClutterRank}, we can provide quantitative predictions of the clutter suppression performance for SF radars. According to the appliaction steps given in subsection \ref{subsubsec:SF}, we have the following corollary.

\begin{corollary}
\label{Co:SF}
(\textit{Stepped-Frequency}) For an SF radar with clutter velocity region $\mathbb{V}$, the clutter rank is
\begin{eqnarray}
\label{Eq:RankSF1}
\mathcal{C}&\approx&\sum_{m\in\mathbb{M}_Q}\min\Big\{\big|\mathcal{I}_m^{\parallel}\{\mathbf{g}\oplus\mathbf{q}\}\big|,\nonumber\\
&&\lceil\frac{2}{c}(f_c+m\Delta{f})\langle\mathbb{V}\rangle\langle T\lfloor\frac {\mathcal{I}_m^{\parallel}\{\mathbf{g}\oplus\mathbf{q}\}}{Q}\rfloor\rangle\rceil+1,\nonumber\\
&&\sum_{\mathcal{T}\in\mathbb{T}_m(\mathbb{V})}\lceil\frac{2}{c}(f_c+m\Delta{f})\langle\mathbb{V}\rangle\langle\mathcal{T}\rangle\rceil+1\Big\}.
\end{eqnarray}
\end{corollary}

\begin{figure}[!t]
\centering
\includegraphics[width=.5\textwidth]{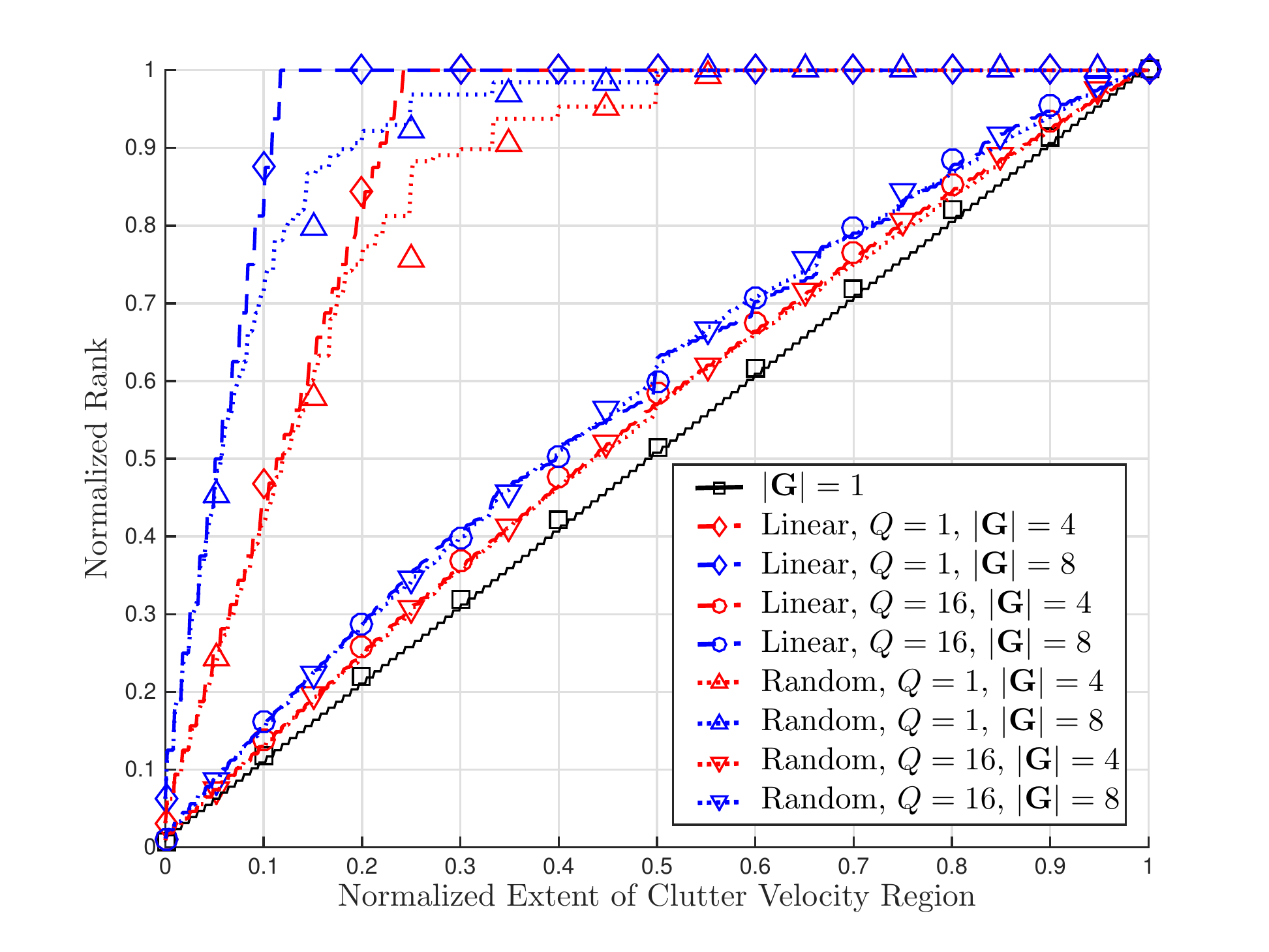}
\caption{NCRs and their approximations for the SF radar pulse trains.}
\label{Fig:Rank_All_SF}
\end{figure}

The NCRs and their approximations for both the LSF and the RSF radars are illustrated in Fig. \ref{Fig:Rank_All_SF}. The pulse number was $256$, and the carrier frequency of each pulse was selected from a set of $|\mathbf{G}|=1$, $4$, and $8$ integers. Both monotone ($Q=1$) and wideband ($Q=16$) pulse waveforms were simulated. Results are shown with legends similar to those in Fig. \ref{Fig:Rank_All_FDA}.

The results revealed by Fig. \ref{Fig:Rank_All_SF} are analogous to Fig. \ref{Fig:Rank_All_FDA}: Fixed-frequency radars have the lowest NCRs; the higher the carrier frequency number, the larger the NCR; the NCRs of wideband pulse waveforms are much smaller than those of monotone pulse waveforms; and comparing to the LSF, the advantages of RSF pulse trains are apparent.
\subsection{FD-MIMO and its STAP Applications}
As introduced in \cite{sammartino2013frequency}, the MIMO technique will remarkably increase the measurement dimension of an FDA radar. Moreover, the results in this section will show that compared to the FDA , the MIMO antenna can also alleviate the increases in NCRs caused by frequency diversity.

Following the steps given in subsection \ref{subsubsec:FD-MIMO}, the RFD-MIMO can be easily applied to FD-MIMO, and the clutter rank can be evaluated as in Corollary \ref{Co:FD-MIMO}.

\begin{corollary}
\label{Co:FD-MIMO}
(\textit{FD-MIMO}) For an FD-MIMO radar with clutter direction region $\mathbb{A}$, the clutter rank is
\begin{eqnarray}
\label{Eq:RankFD-MIMO}
\mathcal{C}&\approx&\sum_{m\in\mathbb{M}_Q}\min\Big\{\big|(d_T\mathcal{I}_m^{=}\{\mathbf{G}_Q\} )\oplus(d_R\mathbf{r} )\big|,\nonumber\\
&& \lceil\frac{1}{c}(f_c+m\Delta{f}))\langle\mathbb{A}\rangle\langle(d_T\mathcal{I}_m^{=}\{\mathbf{G}_Q\} )\oplus(d_R\mathbf{r} )\rangle\rceil+1,\nonumber\\
&&\sum_{\mathcal{S}\in{\mathbb{S}}_m({\mathbb{A}})}\lceil\frac{1}{c}(f_c+m\Delta{f})\langle\mathbb{A}\rangle\langle{\mathcal{S}}\rangle\rceil+1\Big\},
\end{eqnarray}
where $\mathbf{G}_Q$ is formulated as in (\ref{Eq:aFDCM_MIMO}).
\end{corollary}

\begin{figure}[!t]
\centering
\includegraphics[width=.5\textwidth]{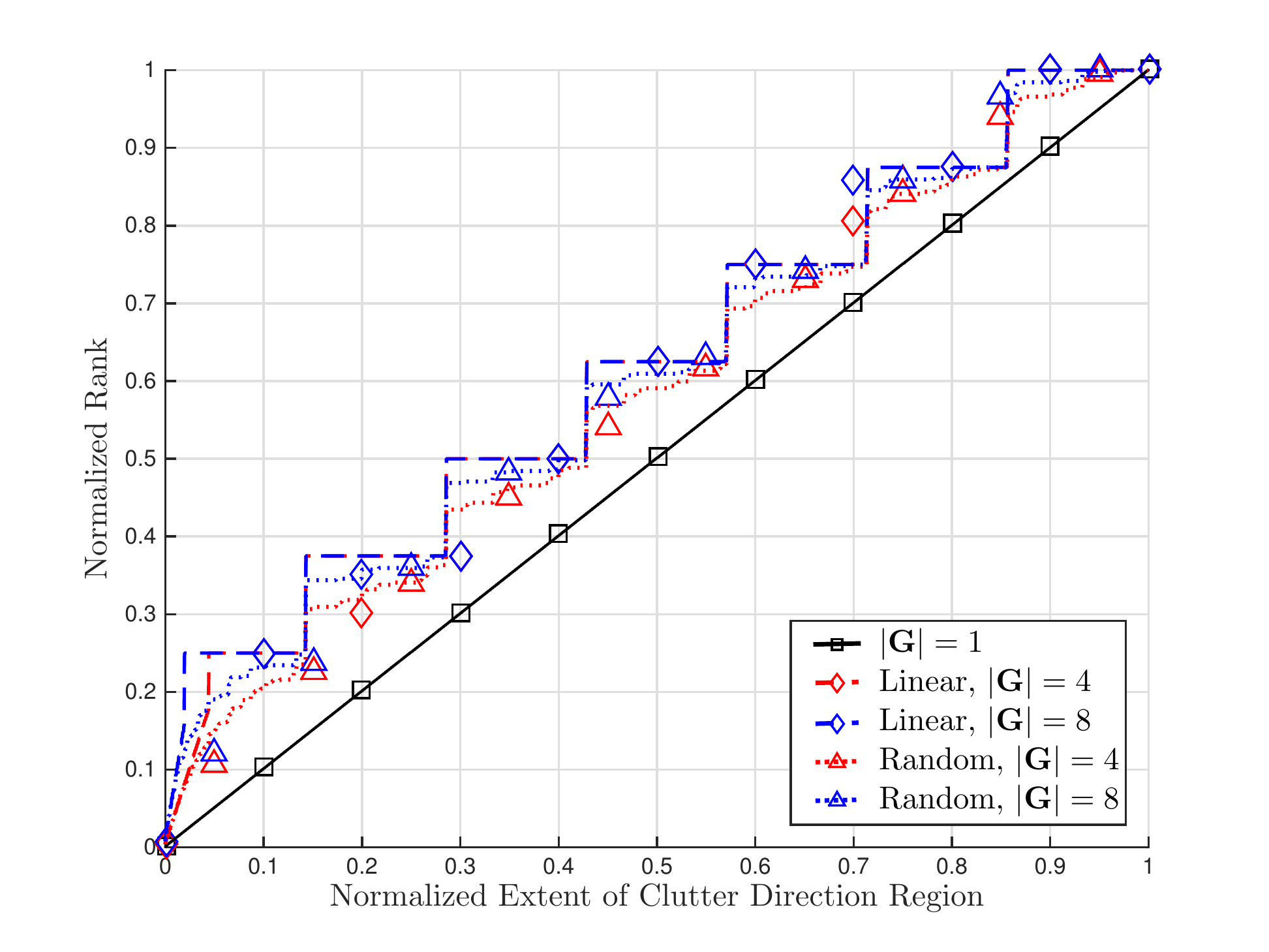}
\caption{NCRs and their approximations for the FD-MIMO radar waveforms.}
\label{Fig:Rank_All_FD_MIMO}
\end{figure}

Simulations of the FD-MIMO radar were conducted with a virtual array of 512 elements, which was synthesized by 64 transmitting and 8 receiving array elements. The distance between the receiving elements was half the wavelength, and $d_T=8d_R$. Similarly, $|\mathbf{G}|=1$, $4$, and $8$ carrier frequencies were respectively simulated with both linear and random carrier frequency assignments. However, only monotone pulses were considered, due to the huge computer memory consumption of the wideband pulse cases. It can be seen in Fig. \ref{Fig:Rank_All_FD_MIMO} that, unlike in the FDA or SF radar, the NCRs and the corresponding approximations changed very little for all the carrier frequency numbers. This result means that the MIMO antenna structure leads to lower NCRs for the FD waveforms, and consequently higher clutter suppression potential. 

For the STAP applications of an airborne FD-MIMO radar, the ground clutter is usually supposed to be stable, and its velocity relative to the radar antenna is caused by the platform's speed. Therefore, the temporal sampling apertures can be embedded into the spatial sampling apertures to form a new group of sampling apertures, given by
\begin{equation}
\big((d_T\mathcal{I}_m^{=}\{\mathbf{G}_Q\} )\oplus(d_R\mathbf{r} )\big)\oplus (2v_pT\mathbf{p}), m\in{\mathbb{M}_Q}.
\end{equation}

Thus the clutter rank of an airborne FD-MIMO with side-looking mode can be evaluated by the following corollary.
\begin{corollary}
\label{Co:FD-MIMO-STAP}
(\textit{FD-MIMO STAP}) For an airborne FD-MIMO radar with side-looking mode, if the beam coverage of the array element is $\mathbb{A}$, the clutter rank is
\begin{eqnarray}
\label{Eq:RankFD-MIMO-STAP}
\mathcal{C}&\approx&\sum_{m\in\mathbb{M}_Q}\min\Big\{\big|\big((d_T\mathcal{I}_m^{=}\{\mathbf{G}_Q\} )\oplus(d_R\mathbf{r} )\big)\oplus (2v_pT\mathbf{p})\big|,\nonumber\\
&& \lceil\frac{1}{c}(f_c+m\Delta{f}))\langle\mathbb{A}\rangle\langle\mathbb{E}_m(\mathbb{O})\rangle\rceil+1,\nonumber\\
&&\sum_{\mathcal{E}\in{\mathbb{E}}_m({\mathbb{A}})}\lceil\frac{1}{c}(f_c+m\Delta{f})\langle\mathbb{A}\rangle\langle{\mathcal{E}}\rangle\rceil+1\Big\},
\end{eqnarray}
where $\mathbb{E}_m(\mathbb{A})$ is the set of sub-apertures defined on the embedded sampling aperture, and $\mathbf{G}_Q$ is formulated as in (\ref{Eq:aFDCM_MIMO}).
\end{corollary}
\begin{figure}[!t]
\centering
\includegraphics[width=.5\textwidth]{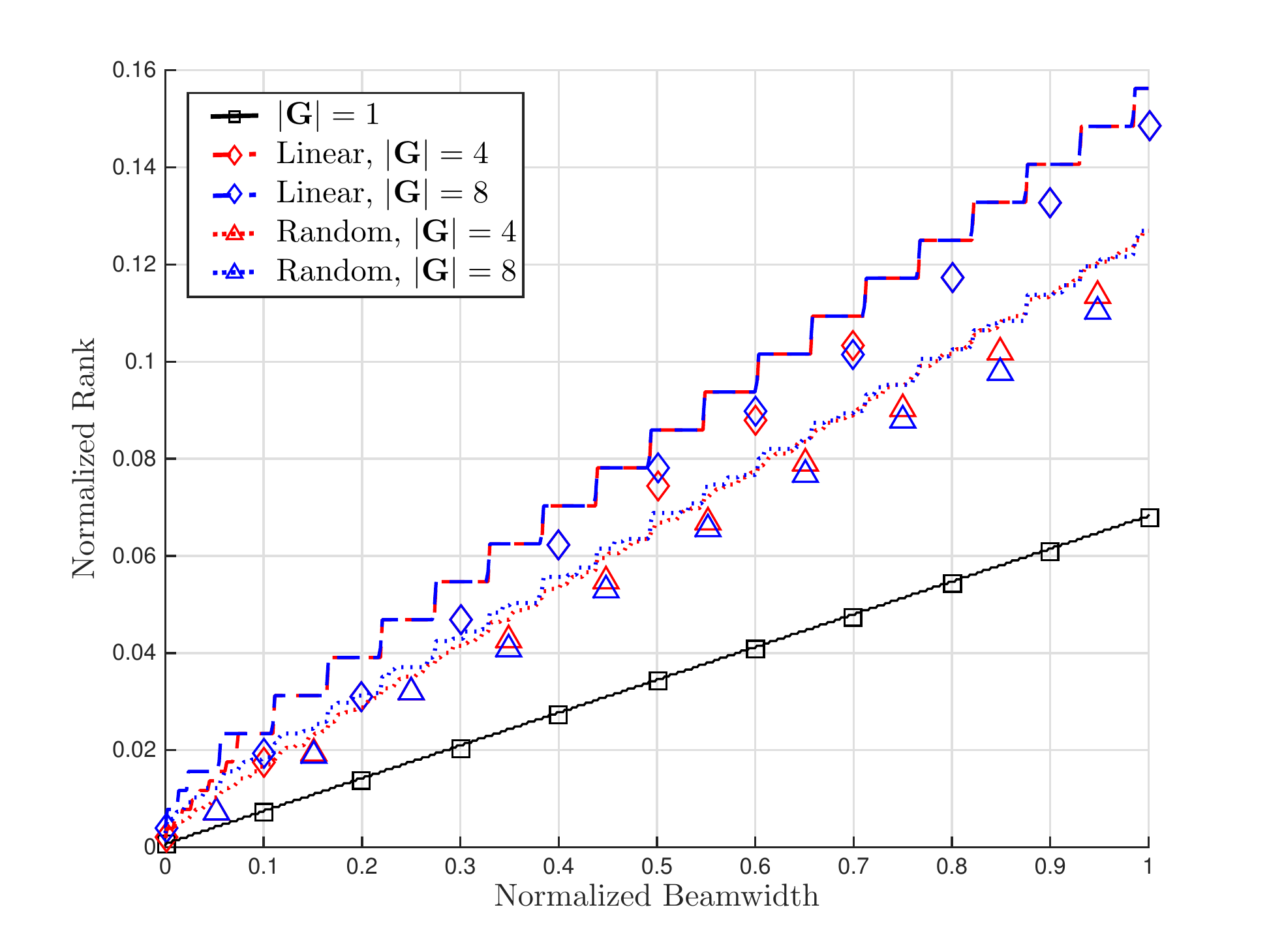}
\caption{The NCRs and their approximations of the airborne FD-MIMO radars with side-looking mode.}
\label{Fig:Rank_All_STAP}
\end{figure}

The simulation results for the FD-MIMO STAP are given in Fig. \ref{Fig:Rank_All_STAP}. The radar had 16 transmitting, and eight receiving array elements, and a pulse train with 16 monotone pulses, which led to a measurement dimension of $PQLR=2048$. A lower NCR than in the above three instance can be expected due to the embedding of the temporal sampling apertures within the spatial ones. As shown in Fig. \ref{Fig:Rank_All_STAP}, for a certain $\langle\mathbb{A}\rangle$, the difference between NCRs of different carrier frequency numbers ($|\mathbf{G}|=4$ and $8$) and different carrier frequency assignments are small, as for the original FD-MIMO waveforms. However, the NCRs of an FD-MIMO STAP are much lower than the other three kinds of FD waveforms, and the FDL is around $0\sim-0.4$ dB for this system configuration.
 
 \subsection{Discussion}
 \label{subsec:discussion}
 
 In this subsection, we give intuitive explanations and further results for the clutter suppression performance of the FDA and SF radars.
 
 As shown in Fig. \ref{Fig:Rank_All_FDA} and Fig. \ref{Fig:Rank_All_SF}, in the FDA or SF radars using monotone pulses, the NCRs of waveforms with linear carrier frequency assignments increase faster than those of their random counterparts. In the FDA or SF radar, the frequency diversity is in a 2D (spatial-spectral or temporal-spectral) domain. In this case, the linear carrier frequency assignment makes the sampling aperture of each frequency point periodic, and the inter-sampling-instant gaps $|\mathbf{G}|$ times larger than those of the fixed-frequency waveforms. Thus when the extents of clutter regions satisfy
 \begin{equation}
 \label{Eq:SplittingTH}
 \langle\mathbb{A}\rangle<\frac{c}{2d_Rf_c|\mathbf{G}|}, \mbox{ or } \langle\mathbb{V}\rangle<\frac{c}{2Tf_c|\mathbf{G}|},\nonumber
 \end{equation}
 aperture splitting will not happen, and
 \begin{equation}
 \begin{array}{ccccc}
\langle\mathcal{S}\rangle&\approx&(L-1)d_T&\mathrm{ where } &|\mathbb{S}_m|=1,\nonumber\\
\langle\mathcal{T}\rangle&\approx&(P-1)T&\mathrm{ where } &|\mathbb{T}_m|=1.\nonumber\\
\end{array}
 \end{equation}
Hence according to Theorem \ref{TH:ClutterRank}, the clutter ranks will grow $|\mathbf{G}|$ times faster than in the fixed-frequency cases, until they touch the ceiling determined by the measurement dimensions.

However, in the spatial or temporal sampling apertures of random carrier frequency assignments, there must be gap(s) between successive sampling instants which is(are) larger than the splitting threshold. The sampling apertures begin dividing into sub-apertures when the extents of clutter regions are small, which makes the clutter ranks grow more moderately, as shown by the triangles and dotted lines in Fig. \ref{Fig:Rank_All_FDA} and Fig. \ref{Fig:Rank_All_SF}.

Through the above analysis, we conclude that in FDA or SF radars with monotone pulses, random frequency assignments can give a lower NCR, and provide a better clutter suppression potential than linear frequency assignments.

However, for a waveform with wideband pulses, if the pulse bandwidth is much larger than the range of the carrier frequencies, then
\begin{equation}
Q\Delta{f}\gg \Delta{f}\langle\mathbf{G}\rangle,\nonumber
\end{equation}
the sub-bands of each pulse will not only increase the measurement dimensions, but also make the the sampling apertures of each frequency point densely and evenly distributed. Thus according to  Theorem \ref{TH:ClutterRank}, the clutter rank will approximately increase at a speed that is direct proportion to the measurement dimension, which keeps the NCRs constant.

\begin{figure}[!t]
\centering
\includegraphics[width=.5\textwidth]{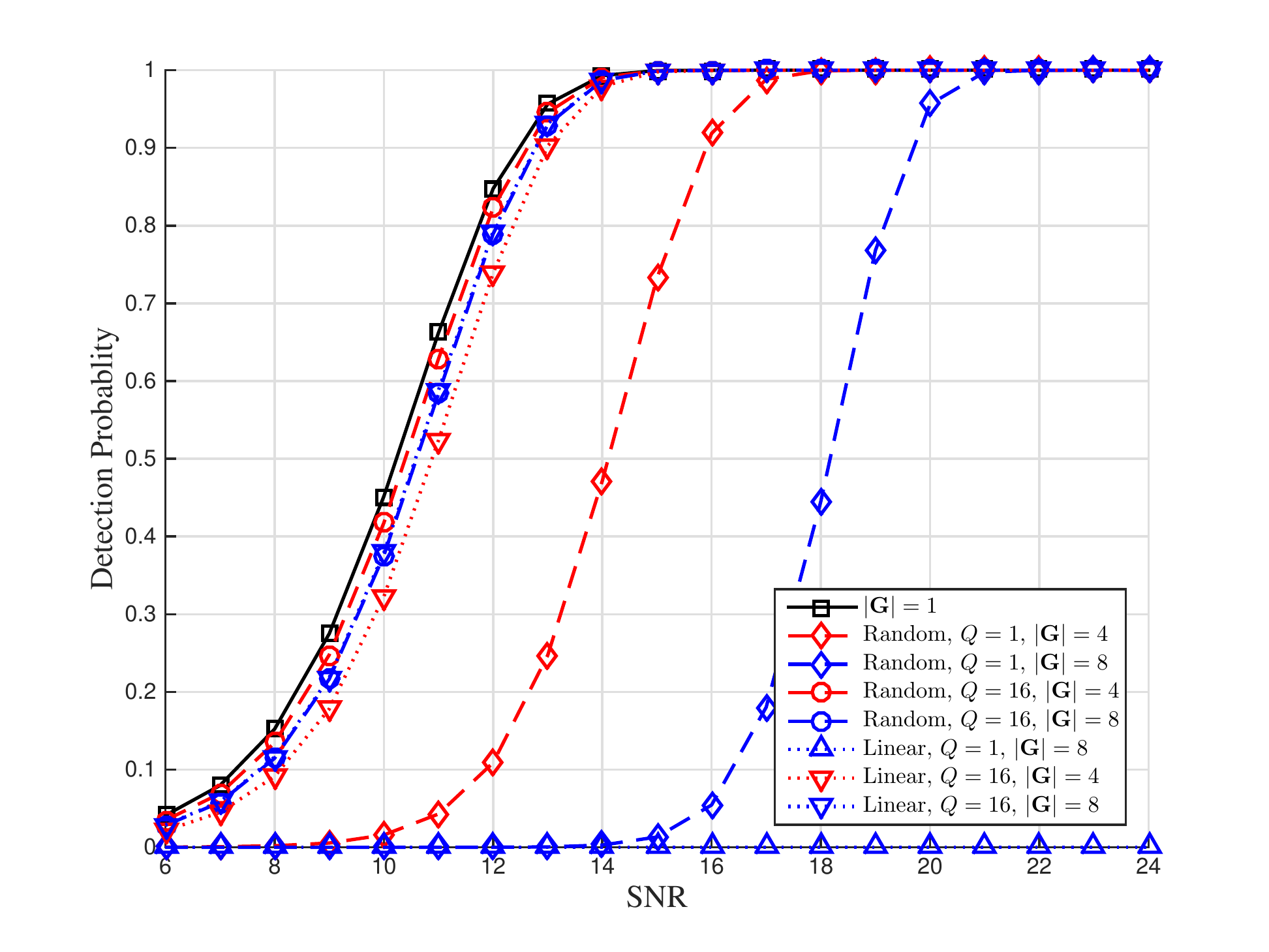}
\caption{Detection probability comparison of FDA radars with different waveform configurations. $P_{\mathrm{fa}}=10^{-5}$.}
\label{Fig:Pd}
\end{figure}

Finally, we provide a  target detection performance comparison between FDA radars with fixed, linearly-assigned, and randomly-assigned carrier frequencies. The waveform parameters used in this simulation were as follows: The number of array elements $L=256$; the inter-element distance $d_R=c/(4f_c)$; the clutter direction region $\mathbb{A}=[-c/(32d_Rf_c),c/(32d_Rf_c)]$; the numbers of carrier frequencies $|\mathbf{G}|=1$, $4$, and $8$; and the pulse bandwidths $Q=1\Delta{f}$ and $Q=16\Delta{f}$. A point target was placed outside the clutter direction region, and the match-filted SNR varied from $6$ dB to $24$ dB. The detection probabilities ($P_{\mathrm{d}}$) of the different waveform configurations were simulated, while the false alarm rates ($P_{\mathrm{fa}}$) were kept constant at $P_{\mathrm{fa}}=10^{-5}$. The results given in Fig. \ref{Fig:Pd} show that the fixed-frequency waveforms have the best detection performance, and the detection probabilities of the wideband pulses are higher than those of the monotone pulses. Moreover, for the RFDA of monotone pulses, the waveforms with a smaller number of carrier frequencies perform better. For the LFDA of monotone pulses, when $|\mathbf{G}=8|$, the clutter spreads over the whole signal space, which makes  $P_{\mathrm{d}}$ always equal to zero. However, the results of $|\mathbf{G}=4|$ were ignored, because the $P_{\mathrm{d}}$ was analogous to the fixed-frequency in some part of the non-clutter region (otherwise zero), which made the averaged detection probabilities pointless.

In addition, the simulation results of SF pulse trains were similar to those of FDA, hence we omitted them.

\section{Conclusion}
\label{sec:Conclusion}
In this paper, we constructed a new FD radar waveform, named RFD-MIMO, by combining two FD waveforms, the FDA and the SF pulse train. The RFD-MIMO can be adopted as a general model, and applied to specific FD waveforms by easy adaptions. Furthermore, by exploring the block diagonal features of the CCM, we proposed an effective approach to estimate the clutter rank of the new waveform model. Then the clutter suppression performances of typical FD waveforms were quantitively evaluated by the corollaries of the approach. Numerical results verified the estimation approach, and revealed two properties of the frequency diversity radars's clutter: The random carrier frequency assignments are more advantageous than their linear counterparts in the coherent clutter suppression of FDA or SF radars, and wideband pulses and MIMO antennas are more suitable for frequency diversity radars to detect targets in clutter.

\appendices
\section{}
\label{AP:One}
For the first inequality in (\ref{Eq:BlockRank}), the integrals in (\ref{Eq:ClutterCovarianceSub}) and (\ref{Eq:ClutterCovarianceSub2}) can be approximated by summations:
\begin{eqnarray}
\label{Eq:ApprocRCm}
\mathbf{R}_{\mathrm{C}_m}&=&\int_{\mathbb{V}}\int_{\mathbb{A}}(\mathbf{u}_{\mathrm{V}_m}\odot\mathbf{u}_{\mathrm{A}_m})(\mathbf{u}_{\mathrm{A}_m}\odot\mathbf{u}_{\mathrm{A}_m})^Hdvd\alpha\nonumber\\
&\approx&(\mathbf{V}^H\circledast\mathbf{A}^H)^H(\mathbf{V}^H\circledast\mathbf{A}^H),
\end{eqnarray}
where $\mathbf{V}=\mathbf{C}^H_{\mathrm{V}_m}$, $\mathbf{A}=\mathbf{C}^H_{\mathrm{A}_m}$, and
\begin{equation}
\label{Eq:KRProduct}
\mathcal{R}\{\mathbf{R}_{\mathrm{C}_m}\}=\mathcal{R}\{\mathbf{V}\circledast\mathbf{A}\}.
\end{equation}

In the case that $\mathcal{R}\{\mathbf{R}_{\mathrm{V}_m}\}+\mathcal{R}\{\mathbf{R}_{\mathrm{A}_m}\}-1\leq K_m$, if
\begin{equation}
\label{Eq:Converse}
\mathcal{R}\{\mathbf{V}\circledast\mathbf{A}\}=\mathcal{R}\{\mathbf{R}_{\mathrm{C}_m}\}<\mathcal{R}\{\mathbf{R}_{\mathrm{V}_m}\}+\mathcal{R}\{\mathbf{R}_{\mathrm{A}_m}\}-1\triangleq U,
\end{equation}
every subset of $U$ columns in $\mathbf{V}\circledast\mathbf{A}$ is linearly dependent. Thus $\exists\mathbf{c}\in\mathbb{C}^{U\times 1}$, which satisfies that
\begin{eqnarray}
\label{Eq:LinearlyDependent}
&&(\Lambda_{\mathbb{U}}\{\mathbf{V}\}\circledast\Lambda_{\mathbb{U}}\{\mathbf{A}\})\cdot\mathbf{c}=\mathbf{0},\\
&&\forall \mathrm{ }\mathbb{U}\subseteq \{0,1,\dots,K_m-1\} \mathrm{, and } |\mathbb{U}|=U,\nonumber
\end{eqnarray}
where $\Lambda_{\mathbb{U}}\{\mathbf{V}\}$ and $\Lambda_{\mathbb{U}}\{\mathbf{A}\}$ are subsets of $U$ columns in $\mathbf{V}$ and $\mathbf{A}$ with a same column index set, $\mathbb{U}$. Because (\ref{Eq:LinearlyDependent}) equals to that $\vectorize\{\Lambda_{\mathbb{U}}\{\mathbf{V}\}(\diag\{\mathbf{c}\}\Lambda_{\mathbb{U}}\{\mathbf{A}\}^H)\}=\mathbf{0}$,
\begin{eqnarray}
0&=&\mathcal{R}\{\Lambda_{\mathbb{U}}\{\mathbf{V}\}(\diag\{\mathbf{c}\}\Lambda_{\mathbb{U}}\{\mathbf{A}\}^H)\}\nonumber\\
&\geq&\mathcal{R}\{\Lambda_{\mathbb{U}}\{\mathbf{V}\}\}+\mathcal{R}\{\Lambda_{\mathbb{U}}\{\mathbf{A}\}\}-U,\nonumber
\end{eqnarray}
which means
\begin{equation}
\label{Eq:RankIneq}
\mathcal{R}\{\Lambda_{\mathbb{U}}\{\mathbf{V}\}\}+\mathcal{R}\{\Lambda_{\mathbb{U}}\{\mathbf{A}\}\}\leq\mathcal{R}\{\mathbf{V}\}+\mathcal{R}\{\mathbf{A}\}-1.
\end{equation}

According to the formulations of the columns in $\mathbf{V}$, each entry of the column index set, $\{0,1,\dots, K_m-1\}$, corresponds to a sampling instant in the temporal sampling aperture. Thus $\mathbf{V}$'s column index set can be divided into two subsets. The first one, termed $\bar{\mathbb{V}}$, corresponds to all the unique sampling instants; the other, termed $\tilde{\mathbb{V}}$,  corresponds to the redundant instants, each of which is a replica of an entry in $\bar{\mathbb{V}}$. Subsets $\bar{\mathbb{A}}$ and $\tilde{\mathbb{A}}$ are defined similarly. Furthermore, from the format properties of the temporal and spatial sampling apertures, we have that\begin{equation}
|\bar{\mathbb{V}} \cap \bar{\mathbb{A}}|\geq 1.
\end{equation}

Moreover, in the approximation in (\ref{Eq:ApprocRCm}), the clutter direction and velocity regions are divided into uniform grids, which makes $\mathbf{V}$ and $\mathbf{A}$ Vandermonde matrices. It has been proven in \cite{sidiropoulos2001identifiability} that, for a Vandermonde matrix, its Kruskal-rank (determined when every subset of Kruskal-rank columns in the matrix is linearly independent and at least one subset of Kruskal-rank$+1$ columns is linearly dependent) equals its rank. Thus every subset of $\mathcal{R}\{\mathbf{V}\}$ columns in $\Lambda_{\bar{\mathbb{V}}}\{\mathbf{V}\}$ and every subset of $\mathcal{R}\{\mathbf{A}\}$ columns in $\Lambda_{\bar{\mathbb{A}}}\{\mathbf{A}\}$ are linearly independent. Then it can be verified that the column index set,
\begin{equation}
\mathbb{U}'\triangleq(\bar{\mathbb{V}} \cap \bar{\mathbb{A}}) \cup \hat{\mathbb{V}} \cup \hat{\mathbb{A}},
\end{equation}
where $\hat{\mathbb{V}}$, $\hat{\mathbb{A}}$ are arbitrary sets, and
\begin{eqnarray}
\hat{\mathbb{V}}\subseteq \bar{\mathbb{V}}\setminus(\bar{\mathbb{V}} \cap \bar{\mathbb{A}}),&|\hat{\mathbb{V}}|=\mathcal{R}\{\mathbf{V}\}-|\bar{\mathbb{V}} \cap \bar{\mathbb{A}}|,\nonumber\\
\hat{\mathbb{A}}\subseteq \bar{\mathbb{A}}\setminus(\bar{\mathbb{V}} \cap \bar{\mathbb{A}}),&|\hat{\mathbb{A}}|=\mathcal{R}\{\mathbf{A}\}-|\bar{\mathbb{V}} \cap \bar{\mathbb{A}}|,\nonumber
\end{eqnarray}
satisfies
\begin{equation}
\label{Eq:Violate1}
|\mathbb{U}'|\leq U,
\end{equation}
and
\begin{equation}
\label{Eq:Violate2}
\begin{array}{ccc}
\mathcal{R}\{\Lambda_{\mathbb{U}'}\{\mathbf{V}\}\}&=&\mathcal{R}\{\mathbf{V}\},\\
\mathcal{R}\{\Lambda_{\mathbb{U}'}\{\mathbf{A}\}\}&=&\mathcal{R}\{\mathbf{A}\}.
\end{array}
\end{equation}

Equation (\ref{Eq:Violate1}) and (\ref{Eq:Violate2}) violate the result in (\ref{Eq:RankIneq}), which is an inference of the assumptions in (\ref{Eq:Converse}). Hence the first inequality in (\ref{Eq:BlockRank}) in proven.

The second inequality in (\ref{Eq:BlockRank}) is straightforward, due to the property that the rank of a Hadamard product of two matrices is no larger than the product of the two matrices' ranks\cite{gentle2007matrix}.

Lemma \ref{LE:RankSupRC} is proven. 

\section{}
\label{AP:Two}

According to the optimal receiver theory \cite{van2004detection}, the optimal linear receiver filtering is the one which achieves the highest output SCNR for a specific waveform and clutter. For a point target with range $D$, velocity $v$, and direction $\alpha$, if the receiver noise is additive white Gaussian with a noise power $\sigma^2$, the optimal clutter suppression performance can be achieved by optimizing the filtered SCNR w.r.t. the filter coefficient vector $\mathbf{w}$:
\begin{equation}
\label{Eq:MaxSCNR}
\mathbf{w}=\arg\max_{\mathbf{w}} \frac{\|\mathbf{w}^H\mathbf{u}(D,v,\alpha)\|^2_2}{\mathbf{w}^H\left(\mathbf{R}_{\mathrm{C}}+\sigma^2\mathbf{I}\right)\mathbf{w}},
\end{equation}
where $\mathbf{w}\in\mathbb{C}^{(PQLR)\times 1}$. 

By minimum variance distortionless response (MVDR) beamforming \cite{van2004detection}, (\ref{Eq:MaxSCNR}) can be solved analytically, where the optimized SCNR is expressed by
\begin{equation}
\label{Eq:MaxSCNRRst}
\mathrm{SCNR}_{\mathrm{opt}}=\mathbf{u}^{H}(D, v,\alpha)(\mathbf{R}_{\mathrm{C}}+\sigma^2\mathbf{I})^{-1}\mathbf{u}(D, v,\alpha).
\end{equation}

In (\ref{Eq:MaxSCNRRst}), the inversion of the matrix $\mathbf{R}_{\mathrm{C}}+\sigma^2\mathbf{I}$ can be calculated by eigen-decomposition. Because the eigenvectors of $\mathbf{R}_{\mathrm{C}}$ are also those of $\mathbf{R}_{\mathrm{C}}+\sigma^2\mathbf{I}$, the eigenspace of $\mathbf{R}_{\mathrm{C}}+\sigma^2\mathbf{I}$ can be divided into the clutter-subspace and the noise-subspace, where the clutter-subspace is spanned by the eigenvectors of $\mathbf{R}_{\mathrm{C}}$, $\{\mathbf{v}_i\}_{i=1}^{\mathcal{C}}$:
\begin{equation}
\label{Eq:EVDCCM}
\mathbf{R}_{\mathrm{C}}=\sum_{i=1}^{\mathcal{C}}\lambda_i\mathbf{v}_i\mathbf{v}_i^H.
\end{equation}
In (\ref{Eq:EVDCCM}), $\{\lambda_i, \mathbf{v}_i\}$ is the $i$th eigenvalue-eigenvector pair of $\mathbf{R}_{\mathrm{C}}$. Moreover, the dimension of noise-subspace is $PQLR-\mathcal{C}$, and its orthogonal bases, termed $\{\mathbf{v}_i\}_{i=\mathcal{C}+1}^{PQLR}$, can be constructed via Gram-Schmidt orthogonalization:
\begin{equation}
\label{Eq:GSofVriance}
\mathbf{R}_{\mathrm{C}}+\sigma^2\mathbf{I}=\sum_{i=1}^{\mathcal{C}}(\lambda_i+\sigma^2)\mathbf{v}_i\mathbf{v}_i^H+\sum_{i=\mathcal{C}+1}^{PQLR}\sigma^2\mathbf{v}_i\mathbf{v}_i^H.
\end{equation}
Substituting (\ref{Eq:GSofVriance}) into (\ref{Eq:MaxSCNRRst}), the maximized output SCNR is
\begin{eqnarray}
\label{Eq:OptSCNRVec}
\mathrm{SCNR}_{\mathrm{opt}}&=&\sum_{i=1}^{\mathcal{C}}\frac{1}{\lambda_i+\sigma^2}\|\mathbf{v}_i^H\mathbf{u}(D,v,\alpha)\|_2^2+\nonumber\\
&&\frac{1}{\sigma^2}\sum_{i=\mathcal{C}+1}^{PQLR}\|\mathbf{v}_i^H\mathbf{u}(D,v,\alpha)\|_2^2.
\end{eqnarray}

In (\ref{Eq:EVDCCM}), $\lambda_i$ equals the eigen-spectral distribution of the clutter power, which can be considered as dominantly large, compared to the noise power in practice\cite{klemm2002principles}. Thus $1/(\lambda_i+\sigma^2)$, the coefficient of the first part in the right side of (\ref{Eq:OptSCNRVec}), approaches zero. Therefore, $\mathrm{SCNR}_{\mathrm{opt}}$ can be approximated by
\begin{eqnarray}
\label{Eq:OptSCNRVecAppro}
\mathrm{SCNR}_{\mathrm{opt}}&\approx& \frac{1}{\sigma^2}\sum_{i=\mathcal{C}+1}^{PQLR}\|\mathbf{v}_i^H\mathbf{u}(D,v,\alpha)\|_2^2\nonumber\\
&=&\frac{1}{\sigma^2}\|\mathbf{P}_{\mathbf{R}_{\mathrm{C}}}^{\perp}\cdot\mathbf{u}(D,v,\alpha)\|_2^2.
\end{eqnarray}

\section*{Acknowledgment}
The authors would like to thank Mr. James Ballard for the proofreading.
\ifCLASSOPTIONcaptionsoff
  \newpage
\fi



%

\bibliographystyle{IEEEtran}
\bibliography{OnClutterRank.bbl}

%
%

%

%
%
%




\end{document}